\documentclass[preprint,12pt]{iopart}

\usepackage{iopams}

\usepackage{graphicx,epsfig}
\usepackage{graphicx}
\usepackage[english]{babel}
\usepackage[colorlinks,ps2pdf]{hyperref}
\usepackage{amsfonts}
\usepackage{amssymb}
\usepackage{array}
\usepackage{pstricks}
\usepackage{color}
\usepackage{xcolor}
\usepackage{subfigure}

\expandafter\let\csname equation*\endcsname\relax
\expandafter\let\csname endequation*\endcsname\relax
\usepackage{amsmath}
\usepackage{latexsym}

\begin{document}

\title[Redshift in a six-dimensional classical Kaluza-Klein type model]{Redshift in a six-dimensional classical Kaluza-Klein type model}

\author{Jacek Syska}

\address{Department of Field Theory and Particle Physics, Institute of Physics,
University of Silesia, Uniwersytecka 4, 40-007 Katowice, Poland}
\ead{jacek.syska@us.edu.pl}

\vspace{10pt}


\begin{abstract}
Multidimensional theories still remain attractive from the point
of view of better understanding fundamental interactions.
In this paper a six-dimen\-sional Kaluza-Klein type
model at the classical, Einstein's gravity formulation is considered.
The static sphe\-rically symmetric solution of
the six-dimen\-sional Einstein equations coupled to the
Klein-Gordon equation with the massless dilatonic field is presented.
As it is horizon free, it is fundamentally different from the four-dimen\-sional Schwarzschild solution. The motion of test particles in such a spherically symmetric configuration is then analyzed. The presence of the dilatonic field has a similar dynamical effect as the
existence of additional massive matter.
The emphasis is put on some observable
quantities like redshifts. It has been suggested that strange features
of emission lines from galactic nuclei as well as quasar-galaxy associations may in fact be manifestations of the
multidimensionality of the world.\\
\\
Keywords: multidimensional theories, Hamilton-Jacobi equation,  redshifts,
quasar-galaxy associations
%
\end{abstract}


\maketitle

\normalsize

\section{\label{noncosm-intr}Introduction}

\hfill{And do not be called teachers;

\hfill{for One is your Teacher,}

\hfill{the Christ {\it Jesus}.}

\vspace{2 mm}

\hfill{Holy Bible, Matthew 23:10}

\vspace{7 mm}

Let the gravitational interaction be described by the Einstein equations \cite{particle-data-group-gravity}, {\it \cite{footnote-1}}, \cite{Denisov-Logunov} and let us suppose that space-time is more than four dimensional \cite{particle-data-group-extra-dim}.
In fact, many 20th century ideas of theoretical physics introduced the  possibility that the world may be multidimensional \cite{Gr-Sch-Kaku}.
Among them are the Kaluza-Klein theories
\cite{Kerner} and despite some problems
the continued interest in them also stems from the
fact that the isometry group of the metric
on the extra--dimensional compact internal space
generates gauge symmetries
of the resulting four--dimensional (non-Abelian) gauge
theories   \cite{Bailin-Love,Choquet-Bruhat}.
%
%
%
In order that a multidimensional theory,
e.g. a ten-dimensional superstring one,
could be taken as the physically accepted one,  it
should possess the proper four-dimensional phenomenology,
primarily the observed bo\-so\-nic and fer\-mio\-nic fields
spectrum \cite{Green-Schwarz-Witten}.
In this context, new results from the large hadron collider (LHC) experiment are still awaited, including the adjustment of all experimental results \cite{babar}. \\
Yet, even if the extra-dimensions effect were noticed
in the LHC experiment, the interpretation of this fact could
possibly possess
the side affect of the (recently increasing) problems,
which lie behind
the experimental validation of the
uncertainty relation (UR) \cite{Generalized-Uncertainty-Principle}.
In that context,
two effects can compete and both of them necessitate a deep theoretical reformulation of UR \cite{Ozawa}.
The first effect is connected with e.g. i)~the
diffraction-interferometric experiments for a photon, whe\-re both the uncertainty relation
and the meaning of the half-widths of a pair of functions (time and frequency) related by the Fourier
transform is examined
\cite{uncertainty-relation-experiment-1,Dziekuje-za-channel}
and ii) an experiment of the successive projective measurements
of two non-commuting neutron spin components
\cite{uncertainty-relation-experiment-2}. In these experiments gravitational effects are not perceived and the
decrease of the right-hand side of the UR, where the Plank constant is, is possibly observed.
%
The second effect is connected with the space-time curvature impact, which makes the value of the uncertainty of the distance in the coordinate basis bigger than the true one (compare Eq.(\ref{row_32})). These mean that the Plank constant
is in practice multiplied by a constant or functional factor
\cite{Generalized-Uncertainty-Principle,uncertainty-relation} and in the observation becomes the effective one
\cite{Spanner-Steck-Akram-Matzkin}.
To sum up, the
problems {\it \cite{footnote-2}}, \cite{Dziekuje-za-skrypt}
with the UR and effectiveness of the Plank constant
could be mistakenly taken as the signal
from the extra dimensions in the accelerator experiments. \\
Meanwhile, there were also attempts in the literature  to seek the effects
of the internal space with a small number of extra dimensions (one or two)
in an astrophysical setting.
Previously, these were pursued by Wesson, Lim, Kalligas, Everitt \cite{Wesson}
for one extra dimension and Biesiada, Ma\'{n}ka,
Syska
\cite{Manka-Syska-a,Dziekuje-Ci-Panie-Jezu-Chryste,LORD-Biesiada-Syska-Manka} for two extra dimensions. This line of thinking is worth
developing in order to gain a better understanding of the
possible visibility of manifestations of e.g. the six-dimensionality of the world, not only in the astrophysical observations but, also in
particle experiments \cite{Dziekuje-za-neutron}.
This would mean that the effects of extra dimensions may well be around us \cite{Kaluza-Klein-dark-matter}
in contrast to standard expectations that extremely high energies are indispensable to probe higher dimensions in search of their existence  \cite{standard-extra-dimensions-in-LHC}.
\\
The contents of the present paper suggest that there is an
intimate relation between the manifestations of the six-dimensionality
of this world and so-called dark matter (not dark energy). The first conjecture
on the existence of dark matter in galaxies came into play when Oort  \cite{Oort-32} in 1932
and Zwicky \cite{Zwicky} in 1933  applied the famous virial
theorem to the vertical motion of the stars in the Milky Way and to the radial velocities of members of the Coma cluster, respectively, or even
earlier in 1915 when {\"O}pik \cite{Opik} evaluated the dynamical
density of matter in the Milky Way in the vicinity of the sun. The
problem was revived and became well established in the seventies when
it was demonstrated \cite{Rubin-Ford-Thonnard,Gates-Gyuk-Turner}
that the rotation curves of spiral galaxies were indicative of the possible presence of a
zero luminosity, dark mass, i.e. unseen
in any part of electromagnetic spectrum (except for the visibility of possible
matter which would be its own antimatter). There have been many
suggestions for candidates for dark matter, from the possibly insufficient impact
of massive compact halo objects (MACHOs) \cite{macho} such as
``brown dwarfs'', ``standard'' neutron stars, ``nomad planets''
etc. at the one extreme to hypothetical elementary particles like massive neutrinos, axions and other weakly interacting massive particles (WIMPs), which till now are either unable to give the amount of mass needed or are in contradiction with other parts of proposed models \cite{contradictions}.  What precisely is dark matter in any detection? Thus, one of the questions is: What, from the scientific point of view, are the inflationary models of all sorts with the persistent lack of observations of their main building blocks, i.e. dark matter and dark energy?
%

The present paper provides a description of the properties of a
certain six-dimensional Kaluza-Klein ty\-pe
model.
Section~\ref{general} contains a description of the model along
with motivations for the choice of six-dimensional space-time (see
also \cite{Dziekuje-za-neutron}). Then, the static
spherically symmetric solutions of the multidimensional Einstein
equations coupled to the
Klein-Gordon equation with
the massless dilatonic field  {\it \cite{footnote-4}}
are derived
\cite{LORD-Biesiada-Syska-Manka,Biesiada-Rudnicki-Syska-1b,Dziekuje-Ci-Panie-Jezu-Chryste}. They are in a sense analogous to the familiar four-dimensional Schwarz\-sch\-ild solution
but fundamentally different i.e. they are horizon free. A more
detailed discussion of the properties of these solutions is
presented in Sections~\ref{general} and \ref{prop}.
Some strictly observable quantities,
such as the redshift formulae, are also briefly discussed.
Section~\ref{motion} is devoted to the analysis of the
motion of test particles ruled by the six-dimensional Hamilton-Jacobi equation.
The application of the obtained background
self-con\-sis\-tent solution \cite{LORD-Biesiada-Syska-Manka}
in the description of the
wave-mechani\-cal structure of e.g. the neutron and its
excited states can be found in \cite{Dziekuje-za-neutron}.
Some possible astrophysical and cosmological observational
consequences of the model are also presented in
Section~\ref{motion}
and in Section~\ref{noncosm-concl}, which also contains concluding remarks and
perspectives. They
refer exclusively to
relatively
tight systems, e.g.
the
vicinity of a galactic nucleus and
a binary galaxy or galaxy-quasar system.
The scale invariance of the model is discussed in Section~\ref{Scale invariance}.
Several more formal, although important, remarks are in the Appendices at the end of this paper. Throughout
Section~\ref{general} the natural units (c = $\hbar$ =1) are used
whereas in
Sections~\ref{prop},~\ref{motion}~and~\ref{noncosm-concl}, which
deal with some more observationally oriented issues,
the velocity of light  is reintroduced
explicitly (the Planck constant is irrelevant for this paper considerations).

\vspace{2mm}

\section{Field equations}

\label{general}

\vspace{3mm}

The simplest extension of the familiar four-dimen\-sional space-time
models are five-dimensional ones, which were previously considered  by
Wesson \cite{Wesson}. In the present paper a six - dimensional model that is more robust
is presented.
The motivations for choosing six-dimensional models by many others  were quite diverse. For example, as late as the years 1984-1986, Nishino, Sezgin, Salam  and Bergshoeff in
\cite{Nishino-Salam-Sezgin-Bergshoeff} suggested that one can  obtain the fermion spectrum by the compactification of the extra two dimensions in a supersymmetric model.
Also,  the six-dimensional models of the Kaluza-Klein theory were  previously investigated by Ivashchuk and Melnikov \cite{Iv-Mel}, Bronnikov and Melnikov \cite{Br-Mel-1995} and by Ma\'{n}ka and Syska \cite{Manka-Syska-a}. \\
In \cite{Dziekuje-za-neutron} the statistical, Fisher
informational reason for the six-dimensionality of space-time was given and
the geometrical properties of the resulting  six-dimensional Kaluza-Klein type
model {\it \cite{footnote-3}},
from the  po\-int of view of its impact on the structure of the neutron and its excited states, were also investi\-gated.
The common point of that model \cite{Dziekuje-za-neutron}
and the one presented below is the basic massless scalar (dilatonic) field obtained simultaneously  with the metric tensor field as the self-consistent solution of the coupled Einstein and Klein-Gordon equations. Then, the obtained metric serves as the background for the equation of motion of the new {\it added object}, which can be a field \cite{Dziekuje-za-neutron} or a classical test particle, immersed in the background metric.
The idea of covering the physical structures that are extremely remote in size by one type of mathematical solution is known by the term of scaling in both theoretical and experimental physics  \cite{scaling}.
In this respect, both models, the current one and the one presented in \cite{Dziekuje-za-neutron} possess the same Kaluza-Klein type self-consistent background solution, which formally can be scaled to  all distances. More on this can be found in \cite{Dziekuje-za-neutron} (see also
Section~\ref{Scale invariance}).
Yet, whereas in \cite{Dziekuje-za-neutron} the {\it added object} is governed by the Klein-Gordon equation solved (not self-consistently) in the mentioned background metric and the solution is a wave-mechanical one, in the present paper the added object is the classical test particle moving in accordance with the Hamilton-Jacobi equation. \\
Thus, let us consider a six-dimensional field theory model comprising the gravitational self field described by a metric tensor, $g_{MN}$,
and a real massless ``basic'' scalar field, $\varphi$. This scalar
field $\varphi$ is a dilatonic field hence, just below, the minus
sign  is present in front of its kinetic energy term
\cite{LORD-Biesiada-Syska-Manka,Biesiada-Rudnicki-Syska-1b}.
In a
standard manner we decompose the action into two parts
\begin{eqnarray}
\label{row_dzialanie-EH-fi}
{\cal S} = {\cal S}_{EH} + {\cal S}_{\varphi} \; ,
\end{eqnarray}
where $S_{EH}$ is the Einstein--Hilbert action
\begin{eqnarray}
\label{row_1}
{\cal S}_{EH} = \int d^{6}x \frac{1}{2 \kappa_{6}} \: \sqrt{- g}
\: {\cal R}
\end{eqnarray}
and $S_{\varphi}$ is the action for a real massless dilatonic
field
\begin{eqnarray}
\label{row_2}
{\cal S}_{\varphi} =  \int  d^{6}x \sqrt{- g} \: {\cal
L}_{\varphi}
= - \int d^{6}x \sqrt{- g} \: \frac{1}{2} \: g_{MN} \:
\partial^{M} \varphi \, \partial^{N} \varphi \;  .
\end{eqnarray}
In Eqs.(\ref{row_1}),(\ref{row_2}), $g = det\; g_{MN}$ denotes the determinant of
the metric tensor, ${\cal R}$ is the curvature scalar of six-dimensional (in general curved) space-time and $\kappa_{6}$
denotes the coupling constant of the six-dimensional theory, which is
analogous to the familiar Newtonian gravity constant. ${\cal
L}_{\varphi}$ is the Lagrangian density for a dilatonic massless
field~$\varphi$.
\\
By extremalizing the action given by
Eqs.(\ref{row_dzialanie-EH-fi})-(\ref{row_2}), we obtain the
Einstein equations
\begin{eqnarray}
\label{row_3}
G_{MN}  = \kappa_{6} \, T_{MN} \; ,
\end{eqnarray}
where $G_{MN} = R_{MN} - \frac{1}{2} \ g_{MN} {\cal R}\,$ is the
Einstein tensor, $R_{MN}$ is the six-dimensional Ricci tensor
and $T_{MN}$ is the energy-momentum tensor of a real
dilatonic field $\varphi$, which is given by
\begin{eqnarray}
\label{row_4}
T^{\; M}_{N} = \partial_{N} \varphi \: \frac{\partial {\cal
L}_{\varphi}} {\partial (\partial_{M} \varphi )} - \delta^{\; M}_{N} {\cal L}_{\varphi} \; .
\end{eqnarray}
Variation of the total action ${\cal S}$ with respect to the field
$\varphi$ gives the Klein-Gordon equation
\begin{eqnarray}
\label{row_5}
\Box \varphi = 0 \;\;  ,
\end{eqnarray}
where
\begin{eqnarray} \label{dalamb}
\Box = - \frac{1}{\sqrt{- g}} \;
\partial_{M} (\sqrt{-g}\; g^{MN}
\partial_{N})
\end{eqnarray}
and $g^{MN}$ is the tensor dual to $g_{MN}$.

Now, we assume that we live in the compactified (which is quite a
reasonable assumption) world, where the six-dimensional
space-time is a topological product of ``our'' curved four-dimensional physical space-time (with the metric $g_{\alpha \omega
}, \; \alpha , \omega = 0,1,2,3 $) and the internal space (with
the metric $g_{he}, \; h,e=5,6 $). Therefore, the metric tensor can
be factorized as follows
\begin{eqnarray}  \label{row_6}
g_{MN} = \begin{pmatrix} g_{\alpha \omega } & 0 \cr 0 & g_{he}
\end{pmatrix} \; .
\end{eqnarray}
The four-dimensional diagonal part is assumed to be that of a spherically symmetric geometry
\begin{eqnarray}
\label{row_7}
g_{\alpha \omega} =
\begin{pmatrix} e^{\nu (r)} & & & \cr & - e^{\mu (r)} & 0 & \cr & 0
& - r^{2} & \cr & & & - r^{2} sin^{2}\Theta \cr \end{pmatrix} ,
\end{eqnarray}
where $\nu (r)$ and $\mu (r)$ are (at this stage) two arbitrary functions. \\
Analogously, we take the two-dimensional internal part to be
\begin{eqnarray}  \label{row_8}
g_{he} = \begin{pmatrix} - \varrho^{2} (r) \: cos^{2} \vartheta &
0 \cr 0 & - \varrho^{2} (r) \end{pmatrix} \; .
\end{eqnarray}
The six-dimensional coordinates $(x^{M})$ are denoted by $(t,
r,\Theta,\Phi ,\vartheta,\varsigma)$, where $t \in [0,\infty )$ is
the usual time coordinate, $r\in [0,\infty ),$ $\Theta \in [0,\pi
] $ and $\Phi \in [0,2 \pi ) $ are familiar three-dimensional
spherical coordinates in the macroscopic space; $\vartheta \in
[-\pi,\pi)$
and $\varsigma \in [0, 2 \pi )$ are coordinates in the internal
two-dimensional parametric space and $\varrho \in (0,\infty )$ is
the ``radius'' of this two-dimensional internal space. We assume
that $\varrho (r) $ is the function of the radius $r$ in our
external three-dimen\-sio\-nal
space {\it \cite{footnote-5}}. \\
The internal space is a 2-dimensional parametric space with
an $r$-dependent parameter $\varrho (r)$, which can be represented as
a surface embedded in the three-dimen\-sio\-nal Euclidean space
\begin{eqnarray}
\label{row_9}
\left\{
\begin{array}{lll}
w^{1} = \varrho (r) \: cos \varsigma \;\; , \;\; \varsigma \in [0, 2 \: \pi )
&  &  \\
w^{2} = \varrho (r) \: sin \varsigma \;\; &  &  \\
w^{3} = \varrho (r) \: sin \vartheta \;\; , \;\; \vartheta \in [- \pi, \pi )
\; .  &  &
\end{array}
\right.
\end{eqnarray}
Now using Eqs.(\ref{row_7})-(\ref{row_8}), we can
calculate the components of the Ricci tensor. The nonvanishing
components are \cite{LORD-Biesiada-Syska-Manka}
\begin{eqnarray}
\label{row_10}
R^{t}_{t} &=& \left( 4 \: \varrho^{2} r \nu^{\prime} + 4 \: r^2
\varrho^{\prime} \varrho \, \nu^{\prime} - r^{2} \varrho^{2}
\mu^{\prime}\nu^{\prime} + r^{2} \varrho^{2} (\nu^{\prime})^{2}
\right. \nonumber \\ &+& \left. 2 \: r^{2} \varrho^{2}
\nu^{\prime\prime} \, \right) (4 \: e^{\mu} r^{2}
\varrho^{2})^{-1}
\end{eqnarray}
\begin{eqnarray}  \label{row_11}
R^{r}_{r} &=& \left(- 4 \: \varrho^{2} r \mu^{\prime} - 4 \: r^2
\varrho^{\prime}\varrho \, \mu^{\prime} - r^{2} \varrho^{2}
\mu^{\prime}\nu^{\prime} +  r^{2} \varrho^{2} (\nu^{\prime})^{2}
\right. \nonumber \\
 &+& \left. 8 \: r^{2} \varrho \varrho^{\prime\prime} + 2
\: r^{2} \varrho^{2} \nu^{\prime\prime} \, \right) \left( 4 \:
e^{\mu} r^{2} \varrho^{2} \right)^{-1}
\end{eqnarray}
\begin{eqnarray}
\label{row_12}
R^{\Theta}_{\Theta} =
R^{\Phi}_{\Phi} &=& \left( - 4 \: e^{\mu} \varrho^{2} + 4 \:
\varrho^{2} + 8 r \varrho \varrho^{\prime} - 2 \: r \varrho^{2}
\mu^{\prime} \right. \nonumber \\
&+& \left. 2 \: r \varrho^{2} \nu^{\prime} \, \right) \left( 4 \:
e^{\mu} r^{2} \varrho^{2} \right)^{-1}
\end{eqnarray}
\begin{eqnarray}  \label{row_14}
R^{\vartheta}_{\vartheta} =
R^{\varsigma}_{\varsigma} &=& \left( 8 \: \varrho \, r
\varrho^{\prime} + 4 \: r^2 (\varrho^{\prime})^{2} - 2 \: r^{2}
\varrho \varrho^{\prime}\mu^{\prime} \right. \nonumber \\
&+& \left. 2 \: r^{2} \varrho \varrho^{\prime}\nu^{\prime} + 4 \:
r^{2} \varrho \varrho^{\prime\prime} \, \right) \left( 4 \:
e^{\mu} r^{2} \varrho^{2} \right)^{-1} \! .
\end{eqnarray}
Let us assume that we are looking for a solution of the Einstein
equations (see Eq.(\ref{row_3})) with the Ricci tensor given by
Eqs.(\ref{row_10})-(\ref{row_14}), with $\nu (r) = \mu(r)$, and
with the following boundary conditions
\begin{eqnarray}  \label{row_16}
\lim_{r\rightarrow \infty } \nu (r) = \lim_{r\rightarrow \infty } \mu (r) = 0
\; ,
\end{eqnarray}
\begin{eqnarray}  \label{row_17}
\lim_{r\rightarrow \infty } \varrho (r) = d = constant \neq 0 \; .
\end{eqnarray}
In other words, we are looking for the solution which at spatial
infinity reproduces the flat external four-dimen\-sio\-nal Minkowski
space-time and static internal parametric space of ``radius'' $d$,
which could be of the order of $10^{-33} \, m$. However, a much higher value of $d$ for the discussed background field configuration, which has, from the point of view of particle physics a
very interesting phenomenology, is also
possible (compare \cite{Dziekuje-za-neutron},  where for the neutron $d \sim 10^{-16} \,  m$).

Now, we make an assumption that the dilatonic field $\varphi $ depends
neither on time $t$ nor on the internal coordinates $\vartheta$ and
$\varsigma$. Because of assumed spherical symmetry of the physical
space-time, it is natural to suppose that the dilatonic field
$\varphi$ is the function of the radius $r$ alone
\begin{eqnarray}
\label{phi sferyczne}
\varphi(x^{M}) = \varphi (r) \; .
\end{eqnarray}
We  also impose a boundary condition for the dilatonic
field $\varphi$
\begin{eqnarray}  \label{row_18}
\lim_{r\rightarrow \infty } \varphi (r) = 0 \; ,
\end{eqnarray}
which supplements boundary conditions (\ref{row_16}) and
(\ref{row_17}) for the metric components.\\
By virtue of Eqs.(\ref{row_4}) and (\ref{row_2}), it is easy to see that
the only nonvanishing components of the energy-momentum tensor are
\begin{eqnarray}  \label{row_19}
\! - T^{r}_{r} = T^{t}_{t} = T^{\Theta}_{\Theta}
= T^{\Phi}_{\Phi} = T^{\vartheta}_{\vartheta} =
T^{\varsigma}_{\varsigma} = \frac{1}{2} \, g^{rr} \, (\partial_{r}
\varphi )^{2}  .
\end{eqnarray}\\
Consequently, it is easy to verify that the solution of the
Einstein equations (\ref{row_3}) is
\begin{eqnarray}  \label{row_20}
\nu (r) = \mu (r) = \ln \left( \frac{r}{r + A} \right)
\end{eqnarray}
\begin{eqnarray}  \label{row_21}
\varrho (r) =d \: \sqrt{\frac{r + A}{r}}
\end{eqnarray}
\begin{eqnarray}  \label{varphi solution}
\varphi (r) = \pm \sqrt{\frac{1}{2 \kappa_{6}}} \:\ln \left( \frac{r}{r + A}
\right) \; .
\end{eqnarray}
Hence, we obtain that the only nonzero component of the Ricci
tensor (see Eqs.(\ref{row_10})-(\ref{row_14})) is $R^{r}_{r}$,
which reads
\begin{eqnarray}  \label{row_23}
R^{r}_{r} = \frac{A^{2}}{2 \: r^{3} (r + A)} \; .
\end{eqnarray}
So the curvature scalar ${\cal R}$ is equal to
\begin{eqnarray}  \label{row_24}
{\cal R} = R^{r}_{r} = \frac{A^{2}}{2 \: r^{3} (r + A)} \; ,
\end{eqnarray}
where $A$ is the real constant with the dimensionality of length,
whose value is to be taken from observations for each particular
system. In the derivation of the above solutions, we have used
Eqs.(\ref{row_10})-(\ref{row_14}), which together with
Eq.(\ref{row_19}) imply that all of the six diagonal Einstein
equations are equal to just
one
\begin{eqnarray}  \label{row_25}
\frac{1}{2} \; {\cal R} = \kappa_{6} \, T^{r}_{r} \; .
\end{eqnarray}
{\footnotesize
{\bf Remark}:
\label{f-5}
{\it Putting
Eqs.(\ref{row_25}),(\ref{row_19})~and~(\ref{row_24}) together we
can notice a similarity between the equation
\begin{eqnarray}
\label{screen-in-gravity-cond}
R^{r}_{r} = -\kappa_{6} (\partial_{r} \varphi)^{2} g^{rr} \;\;\;\;\;\;\;\;
\end{eqnarray}
and its electromagnetic analog
$\nabla^2 {\bf A} = m_{A}^{2} {\bf A}$,
where ${\bf A}$ is the electromagnetic vector potential. Eq.(\ref{screen-in-gravity-cond})
is the (anti)screening current condition in gravitation that is analogous to
that in electromagnetism or in the
electroweak sector in the self-consistent approach}
\cite{Dziekuje-Ci-Panie-Jezu-Chryste,Dziekuje-Jacek-nova-2,Dziekuje-za-self-consistent-state}.
} \\
\\
Now, we can rewrite the metric tensor in the form
\begin{eqnarray}
\label{row_26}
\!\!\!\!\!\!\!\!\!\!\!\!\!\!\!\!\!\!\!\!
\!\!\!\!\!\!\!\!\!
g_{MN} &=& {\rm diag} \left( g_{tt}, g_{rr}, g_{\Theta\Theta}, g_{\Phi\Phi}, g_{\vartheta\vartheta}, g_{\varsigma\varsigma} \right)  = \nonumber \\
\!\!\!\!\!\!\!\!\!\!\!\!\!\!\!\!\!\!\!\!
\!\!\!\!\!\!\!\!\!
& = & {\rm diag} \left( \frac{r}{r + A}, \, - \, \frac{r}{r +
A}, \, - \, r^{2}, \, - \, r^{2} sin^{2}\Theta, \,
- d^{2} \frac{r + A}{r} \, cos^{2}\vartheta, \, - \, d^{2} \frac{r
+ A}{r} \right)
\end{eqnarray}
with its  determinant equal to
\begin{eqnarray}
\label{row_27}
g = det g_{MN} = - (d^{2} \: r^{2} \: sin\Theta \: cos\vartheta )^{2} \;
\end{eqnarray}
that itself is nonsingular.
We see that the space-time of
the model is stationary.

It is also necessary to verify whether the solution of the
Klein-Gordon equation (\ref{row_5}) is in agreement with
Eq.(\ref{varphi solution}), which follows from the Einstein
equations. According to Eq.(\ref{row_5}) and Eqs.(\ref{phi sferyczne}),(\ref{row_26}),(\ref{row_27}), we obtain that
\begin{eqnarray}
\label{row_28}
\partial_{r} \varphi (r) = - \, {\rm C} \, g_{rr} \, r^{-2} = {\rm C} \, \frac{1}{r(r + A)}
\; ,
\end{eqnarray}
where ${\rm C}$ is a constant. Comparing this result with Eq.(\ref{varphi solution}), we conclude that if
\begin{eqnarray}
\label{row_29}
{\rm C} = \pm \, \frac{A}{\sqrt{2 \kappa_{6}}}
\end{eqnarray}
then the solution of the Klein-Gordon equation is in agreement
with the solution of the Einstein equations coupled to the
Klein-Gordon one.
Hence, the real massless ``basic''
dilatonic field $\varphi (r)$ given by Eq.(\ref{varphi solution}) can be
the source of the nonzero metric tensor given by Eq.(\ref{row_26}).
Only when the constant $A$ is equal to zero do the solutions
(\ref{row_20})-(\ref{varphi solution}) become trivial and the
six-dimensional space-time is Ricci flat.

It is worth noting that because the components
$R^{\vartheta}_{\vartheta}$ and $R^{\varsigma}_{\varsigma}$ of the
Ricci tensor are equal to zero for all values of $A$, the internal
space is always Ricci flat. However, we must not neglect the
internal parametric space because its ``radius'' $\varrho$ is a
function of $r$ and the two spaces, external and internal, are
therefore ``coupled''. Only when $A=0$ are these two spaces
``decoupled'' and our four-dimensional space-time becomes
Minkowski flat. \\

\hspace{-7mm}
{\footnotesize
{\bf Remark}:
\label{f-6}
{\it
When $A$ is not equal to zero, our
four-dimensional external space-time is curved. Its scalar
curvature ${\cal R}_{4}$ is equal to (\ref{row_24})
}
\begin{eqnarray}  \label{row_30}
{\cal R}_{4} = {\cal R} = \frac{A^{2}}{2 \: r^{3} (r + A)} \; .
\quad\quad\quad\quad\quad\quad
\end{eqnarray}
}

\subsection{The stability of the background
solution}

\label{stability of the background}

Let us consider the stability of the
self-consistent gravito-dilatonic configuration given by
Eqs. (\ref{varphi solution}) and (\ref{row_26}). We calculate its energy $E_{g + \varphi}$ (see \cite{Dziekuje-za-neutron}), which is the integral over the spacelike hypersurface \cite{Choquet-Bruhat}
\begin{eqnarray}
\label{config energy} E_{g + \varphi} = \int_{V} d^{5} x \ \sqrt{-
g} \ \left( G^{tt} + \kappa_{6} \ T^{tt} \right) \; ,
\end{eqnarray}
where using Eqs.(\ref{row_19})-(\ref{row_27}) we obtain
\begin{eqnarray}
\label{conig energy value}
E_{g + \varphi} = - 2 \int_{V} d^{5} x
\ \sqrt{- g} \, g^{tt} \, \frac{\cal R}{2} = - 2 \, Q \leq 0 \;
\end{eqnarray}
and
\begin{eqnarray}
\label{K}
\!\! Q = 8\,d^2\,{\pi }^2 \! \! \lim_{\varepsilon \rightarrow
0^{+}} \! \int_{\varepsilon}^{\infty} \! \frac{A^2}{r^2}\,dr
= (2 {\pi }\, d) \, ( 4\,{\pi }\, A^{2}) \!\! \lim_{\varepsilon
\rightarrow 0^{+}} \! \frac{d}{\varepsilon}   \, .
\end{eqnarray}
By using the Fisher information formalism, which was developed for the physical models by Frieden and others
\cite{Frieden,Dziekuje-informacja,Dziekuje-za-skrypt,Dziekuje-za-neutron,Dziekuje-za-EPR-Bohm}, the foundation of
the partition of $E_{g + \varphi}$ in
Eq.(\ref{config energy}) into two parts, was constructed.
The first term, $G^{tt}$, is connected with the Fisherian kinematical degrees of freedom of
the gravitational configuration and the second one, $T^{tt}$, is connected with its structural degrees of freedom.\\
The integrand
in Eq.(\ref{K}) does not converge  on $(0, \infty)$ for $A \neq 0$.
If in Eq.(\ref{K}), the cutoff $\varepsilon = d$ is
taken \cite{Dziekuje-za-neutron}, which breaks the self-consistency of the solution,
then it leads to $Q = Q_{cut} = (2 {\pi }\, d) \, (4\,{\pi }\,
A^{2})$ making $Q_{cut}$ finite. The value of $Q_{cut}$ could be
small in contradistinction to the infinite value obtained in
Eq.(\ref{K}) for $\varepsilon \rightarrow 0$ taken in the self-consistent case.
Hence, for $A \neq 0$, the self-consistent solution given by
Eqs.(\ref{varphi solution}) and (\ref{row_26}), which has
energy $E_{g + \varphi}$ given by Eq.(\ref{conig energy value}),
cannot be destabilized to yield any other with finite energy.
The energy is the generic property of the solution \cite{Peixoto}.
In particular, if $\psi(B)$ is a
well-behaved field \cite{Rendall} that depends on
a parameter $B$, then it cannot be destabilized to any
solution with the energy $E_{g + \psi(B)}$ from the sequence that has
the limit $E_{g + \psi}$ $ \stackrel{B \rightarrow 0}{\longrightarrow} 0$,
which itself corresponds to $A = 0$.
Hence, the transition from the self-consistent gravito-dilatonic
configuration given by Eqs.(\ref{varphi solution}) and
(\ref{row_26}) to any others with finite energy
is forbidden \cite{Dziekuje-za-neutron}.
%
\\
Let us notice that the solution for $A=0$ is consistent with the
Minkowskian space-time one, which has
global energy equal to zero \cite{Choquet-Bruhat}.
Therefore, the condition on the right hand side of
Eq.(\ref{conig energy value}) indicates that the
self-consistent gravito-dila\-to\-nic configuration
given by Eqs. (\ref{varphi solution}) and (\ref{row_26})
for $A \neq 0$ is more stable than the empty Minkowskian
space-time solution.
Thus, it is worth noting that
this configuration of fields might serve,
under further specific conditions chosen for the particular
physical system
\cite{Dziekuje-za-neutron}, as the background one. It
appears in equations of motions of all new fields that
enter into the system weakly \cite{Dziekuje-za-neutron}.
This suggests that configurations similar to the obtained
self-consis\-tent gravito-dila\-to\-nic one could be the main
building materials
for the observed structures in the universe,
both on the microscopic \cite{Dziekuje-za-neutron} and,
as it is suggested below, on the astrophysical and
cosmological scales, too.

\subsection{The problems of the space-time singularity and Kaluza-Klein type excitations}

\label{K-K problems}

Firstly, in Section~\ref{Radial trajectories} as far as the space-time geometry is discussed, it will be proven  that
the space-time singularity connected with the metric tensor (\ref{row_26}) is the one through which  only the null geodesics could pass through.
Hence, the obtained solution is not in contradiction with
``cosmic censorship'',
a conjecture that is still not proven
\cite{Rendall}. \\
Secondly, the problem to discuss is the Kaluza-Klein type
excitations. In general, in the Kaluza-Klein gravity, the ground
state solutions can contain a number of the resulting four-dimensional (massless or not) fields \cite{Manka-Syska-a}. \\
For example, into the presented model, in addition to the six-dimensional dilatonic
field given by Eq.(\ref{varphi solution}), two massless scalars
tied with the shape
of a two-dimensional internal parametric space (\ref{row_9})
could also be incorporated  \cite{Dziekuje-za-neutron}.
Then, their existence is
highly constrained by observations that usually tend to
rule out the models. However, this is not the fate of our
non-homo\-genous solution.
Indeed,
it was noticed in \cite{Dziekuje-za-neutron}  that
unless additional fields enter, on the level of the equations of motion (\ref{row_3}) and (\ref{row_5}), the theory is scale invariant (see also Section~\ref{Scale invariance}).
Yet, when a new scalar field $\phi$ is introduced then the massless dilatonic field $\varphi$ of the model, which is the Goldstone one,
gives masses to the ``undesirable'' Kaluza-Klein type mo\-du\-li fields.
As the Goldstone mode, the dilatonic field
$\varphi$ is absorbed by the new scalar field $\phi$
introduced into the system \cite{Dziekuje-za-neutron}. As a result, it was proven that the
obtained ground state solution of the total gravito-dilatonic and $\phi$
fields configuration
acquires two spinorial degrees of freedom, where the origin of its
non-zero spin
is perceived as a manifestation of both the geometry
of the internal two-dimensional parametric space (\ref{row_9}) and of the kinematics of the field $\phi$ inside it. The obtained configuration  can be interpreted as, e.g. the neutron
(but other particle solutions are possible also) \cite{Dziekuje-za-neutron}. Therefore, the model can be extended in such
a way that the intrinsic geometrical and kinematical properties of
fields in the extra dimensions
also manifest themselves in
possible observational consequences in the realm of the
physics of one particle.
In consequence, e.g.
the relevant
observable neutron excited states were  also calculated
\cite{Dziekuje-za-neutron}. \\
To secure the stability of this composite model of the neutron, it was found that the size of the parametric space
has to be equal to $d \approx 0.2071 \; {\rm fm}$ \cite{Dziekuje-za-neutron}. This signifies the large extra dimension solution of $d$,
which is of the substantial fraction of the nucleon size, which is
evidently not excluded by the experiment \cite{particle-data-group-extra-dim}.
Simultaneously, the inclusion of the additional sca\-lar field
$\phi$ does not lead to the destruction of our background
dilatonic solution (\ref{varphi solution}) as was argued in  \cite{Dziekuje-za-neutron}. \\
Finally, the problem of quantum fluctuations can also be considered. Yet, the existing quantum field theory (QFT)
interpretation of all physical phenomena could and should be
questioned {\it \cite{footnote-6}}
(see \cite{bib-B-K-1,Dziekuje-Jacek-nova-2,Manka-Syska-a}),
i.e. there may
exist physical realities for which the existence of the quantum
fluctuations is not necessary at all. Nowhere is this problem so
crucial as in the case of gravitational interaction, at least
as far as the experiment is considered. If this is a problem of
general relativity, hence, it is also of its Kaluza-Klein gravity
extensions.

In the present paper the astrophysical and
cosmological significance of the solution is discussed.

\section{\label{prop}Some properties of solutions}

If the parameter $A$ is strictly positive, $A > 0,$
then Eqs.(\ref{row_20})-(\ref{varphi solution}) are valid
for all $r > 0$.
({\it The discussion
of $A<0$ case will be left for
\label{f-7}
\ref{noncosm-app}.})
The metric tensor becomes singular only at $r =
0$; nevertheless, its determinant $g$ (see Eq.(\ref{row_27}))
remains well defined. Below, several formulae
which will be useful in
the later discussion are collected. We start with time
and the radial components of the metric $g_{MN}$ and the internal
``radius'' $\varrho (r)$ (see
Eqs.(\ref{row_21})~and~(\ref{row_26}))
\begin{eqnarray}  \label{row_31}
\left\{
\begin{array}{lll}
g_{tt} = \frac{r}{r + A} \; &  &  \\
g_{rr} = - \frac{r}{r + A} \; &  &  \\
\varrho (r) = d \sqrt{\frac{r + A}{r}} \; \; . &  &
\end{array}
\right.
\end{eqnarray}
It will be also useful to write the explicit relation for the real, physical radial distance $r_{l}$ from the center
\begin{eqnarray}
\label{row_32}
\!\!\!\!\!\!\!\!\!\!\!\!\!\!\!\!\!\!\!\!\!\!\!\!\!\!\!\!\!\!
r_{l}  =  \int_{0}^{r} \!\! dr
\, \sqrt{- g_{rr}} = {\sqrt{{\frac{r}{{r + A}}}}}\,\left( r + A
\right)
+ {\frac{1}{2} \,A\,\ln ({\frac{A}{{A +
2\,r + 2\, (r + A) \,{\sqrt{{\frac{r}{{r + A}}} \, }} }} })} < r  \; .
\end{eqnarray}
%
\begin{table}[tbp]
\label{prop-table}
\begin{center}
{\begin{tabular}{lll}
\hline\noalign{\smallskip}
Example & \!\!\! $M \;  [M_{\odot}]$ & $\; A \;  [pc]$  \\
\noalign{\smallskip}\hline\noalign{\smallskip}
sun & \!\!\! $\; 1.$ & $\; 0.96 \; 10^{-13}$ \\
globular cluster & \!\!\! $\; 10^4 - 10^6$ & $ \; 0.96 \; (10^{-9} - 10^{-7}) $ \\
galactic nucleus$^{*}$ & \!\!\! $\; 10^7$ & $ \; 0.96 \; 10^{-6}$ \\
galaxy & \!\!\! $\; 5. \; 10^{11}$ & $ \; 4.79 \; 10^{-2}$ \\
{\small binary galaxy system}$^{*}$
 & \!\!\! $\; 1. 10^{15}$ & $ \; 0.96 \; 10^{2}$ \\
{\small galaxy--quasar system}$^{*}$
 & \!\!\! $\; 2. \; 10^{18}$ & $ \; 1.91 \; 10^5$ \\
\noalign{\smallskip}\hline
\end{tabular} }
\end{center}
\vspace{-4mm}
\caption{
Values of the parameter $A$ for which the
six-dimensionality of the world influences the dynamics of test
particles in a similar way as the existence (in the 4-dimensional
world) of mass $M$ (given for some examples motivated by
astrophysics). The examples marked by $^{*}$ are
exceptional in the sense that parameter $A$ has been estimated
due to the demand to explain the observed redshift peculiarities of such systems. Hence, the mass $M$ has a purely effective meaning
here --- for details see Sections~\ref{motion}~and~\ref{noncosm-concl}.
}
\end{table}
{\bf Note:} Let us recall that in the standard derivation of the Schwarz\-schild
solution, the free parameter in the metric tensor is identified
with the total mass of a spherically symmetric configuration due to the demand that at large distances the metric tensor should reproduce
the Newtonian potential.
Because $g_{tt} \rightarrow 1 $ for $r \rightarrow \infty $ (see
Eq.(\ref{row_31})), it is interesting to compare the gravitational
potential $g_{tt} = \frac{r}{r + A} \approx 1 - \frac{A}{r}$ for
$r\gg A$ with the gravitational potential $g_{tt} = 1 -
\frac{G}{c^{2}} \frac{2 M}{r}$, which is induced by a mass $M$ in the
Newtonian limit. $G$ and $c$ are the four-dimensional
gravitational constant and the velocity of light, respectively.
Comparing these two potentials, we obtain that $A = 2
\frac{G}{c^{2} } M $, so parameter $A$ can have similar
dynamical consequences as the mass $M$.
({\it The dynamical interpretations of
$A$ are different for other powers of
\label{f-8}
$\frac{A}{r}$.})
Table~1
contains some
astrophysically interesting masses that mimic the values of
parameter $A$.
In this case, the gravitational potential $g_{tt}$ (see
Eq.(\ref{row_26}) or Eq.(\ref{row_31})) is attractive although
there is no massive matter acting as a source.
Nevertheless, we cannot essentially
identify parameter $A$ directly with $M$. The reason is that
the presented solution
describes a case where ordinary matter is absent and the (only)
contribution to the energy-momentum tensor comes from the
infinitely stretched, massless dilatonic field $\varphi.$

It is well known \cite{Land-Lif} that the
frequency $\omega_{0}$ of light, moving along the geodesic line in a
gravitational field that is static or stationary and measured in
units of time $t$, is constant ($\omega_{0} = constant$) along the
geodesic. The frequency $\omega$ of light as a function of the
proper time~$\tau$ ($d \tau = \sqrt{g_{tt}} \: dt$ ) is equal to
\begin{eqnarray}  \label{row_43}
\omega = \omega_{0} \: \frac{dt}{d\tau} = \frac{\omega_{0}}{\sqrt{g_{tt}}} =
\omega_{0} \: \sqrt{g^{tt}} \; \; .
\end{eqnarray}\\
Let us assume that a photon with frequency $\omega_{\sigma}$ (measured in
units of the proper time $\tau$) is emitted from a source that
is located at a point $r = r_{\sigma}$ where $g_{tt} = g^{\sigma}_{tt}$.
Then the photon is moving along a geodesic line that reaches the
observer ($obs$) at the point $r = r_{obs}$, where $g_{tt} =
g^{obs}_{tt}$, with the frequency $\omega_{obs} $
\begin{eqnarray}  \label{row_44}
\frac{\omega_{obs}}{\omega_{\sigma}} = \sqrt{\frac{g_{obs}^{tt}}{g_{\sigma}^{tt}}} =
\sqrt{\frac{g^{\sigma}_{tt}}{g^{obs}_{tt}}} \; \; .
\end{eqnarray}
Using Eq.(\ref{row_31}) we can rewrite Eq.(\ref{row_44}) as
\begin{eqnarray}  \label{row_45}
\frac{\omega_{obs}}{\omega_{\sigma}} = \frac{\sqrt{\frac{r_{\sigma}}{r_{\sigma} + A}}}{
\sqrt{\frac{r_{obs}}{r_{obs} + A}}} \;.
\end{eqnarray}
For simplicity, let us consider a limiting case in which the observer is
situated at infinity. Then we get
\begin{eqnarray}  \label{row_46}
\frac{\omega_{obs}}{\omega_{\sigma}} = \sqrt{\frac{r^{w}_{\sigma}}{r^{w}_{\sigma} + 1}} \;
\; ,  \;\;\;\; {\rm where} \; \;\;\; r^{w}_{\sigma} = \frac{r_{\sigma}}{A} \; \; .
\end{eqnarray}
Therefore, we see that the nearer the source is to the center
of the field $\varphi (r)$ the more  the emitted photon is
redshifted at the point where it reaches the observer. It should
also be emphasized that this redshift does depend on the relative
radius $r^{w} = \frac{r}{A} $ rather than separately on $r$ and
$A$ (see also
Section~\ref{Scale invariance}).

\section{Six-dimensional Hamilton-Jacobi equation for a ``test particle''}

\label{motion}

In order to gain a better understanding concerning possible
manifestations of the six-dimen\-sionality of the world, let us
investigate the motion of a ``test particle''
of mass $m$
in the central gravitational field described by
Eq.(\ref{row_26}). Whether the moving object can be treated as a
``test particle'' depends on the value of parameter $A$ in the
metric tensor $g_{MN}$ (see Eq.(\ref{row_26})). \\
The Hamilton-Jacobi equation
\begin{eqnarray}
\label{row_49}
g^{MN} \: \partial_{M} S \, \partial_{N} S - m^{2} c^{2} = 0
\end{eqnarray}
describing the motion of a test particle reads
\begin{eqnarray}
\label{row_50}
\!\!\!\!\!\!\!\!\!\!\!\!\!\!\!\!\!\! & & \frac{r + A}{r}\left( \frac{\partial S}{c
\partial t} \right)^{2} - \frac{r + A}{r} \: \left( \frac{\partial
S}{\partial r} \right)^{2} - \frac{1}{r^{2}} \: \left(
\frac{\partial S}{\partial \Theta} \right)^{2} - \frac{1}{r^{2} \: sin^{2} \Theta} \: \left( \frac{\partial S}{\partial \Phi} \right)^{2} \nonumber
\\
\!\!\!\!\!\!\!\!\!\!\!\!\!\!\!\!\!\!
&-&  \frac{1}{d^{2}} \: \frac{r}{r +
A} \: \frac{1}{cos^{2} \vartheta} \: \left( \frac{\partial
S}{\partial \vartheta} \right)^{2}
- \frac{1}{d^{2}} \: \frac{r}{r + A} \: \left( \frac{\partial
S}{\partial \varsigma} \right)^{2} - m^{2} \: c^{2} = 0 \; ,
\end{eqnarray}
where $S$ denotes the complete integral of the Hamilton-Jacobi equation and  $m$ is the
mass of a ``test particle'' in  six-dimensional space-time.

Without loss of generality, we shall restrict ourselves to the motion in the plane $\Theta = {\pi}/2$, thus
\begin{eqnarray}  \label{row_51}
\frac{\partial S}{\partial \Theta} = 0 \; \; .
\end{eqnarray}\\
If the integral $S$ depends on a non-additive constant, which is the energy
${\cal E}_{0}$
(the Hamiltonian does not depend explicitly on time),
then the standard procedure of the separation of variables begins with the following factorization of $S$
\begin{eqnarray}
\label{row_52}
S = - {\cal E}_{0} \, t + {\cal M}_{\Phi} \, \Phi + S_{r} (r) + {\cal M}_{\varsigma} \, \varsigma + S_{\vartheta} (\vartheta)\; .
\end{eqnarray}
The separation constants ${\cal E}_{0}$, ${\cal M}_{\Phi}$ and ${\cal
M}_{\varsigma}$ have the meaning of the total
energy, the effective angular momentum and the internal angular
momentum, respectively.
Then by separating Eq.(\ref{row_50}) into four-dimensional and
internal parts, one arrives at the formula
\begin{eqnarray}  \label{separation}
\!\!\!\!\!\!\!\!\!\!\!\!\!\!\!\!\!\!
& & \frac{r + A}{r} \: \left[ \frac{r + A}{r} \:
(\frac{{\cal E}_{0}}{c})^{2} -  \frac{r + A}{r} \;
(\frac{\partial S_{r}}{\partial r})^{2}
- \frac{1}{r^{2} \: sin^{2} \Theta} \: {\cal M}_{\Phi}^{2} - m^{2} \: c^{2} \right]
\\
\!\!\!\!\!\!\!\!\!\!\!\!\!\!\!\!\!\!
& = & \frac{1}{d^{2}} \: \frac{1}{cos^{2} \vartheta} \: (
\frac{\partial S_{\vartheta}}{\partial \vartheta})^{2} +
\frac{1}{d^{2}} \: {\cal M} _{\varsigma}^{2} =: k_{\vartheta
\varsigma}^{2} = constant \: . \nonumber
\end{eqnarray}
The last equation in (\ref{separation}) is easy to integrate for
$S_{\vartheta}$
\begin{eqnarray}  \label{row_53}
S_{\vartheta} = {\pm} \: ( d^{2} \: k_{\vartheta
\varsigma}^{2} - {\cal M} _{\varsigma}^{2} )^{\frac{1}{2}} \: sin
\vartheta =: {\pm} \: k_{\vartheta} \, d \: sin \vartheta \; ,
\end{eqnarray}
where $k_{\vartheta}$ has the meaning of the internal momentum
and one can distinguish the square of the (total) internal momentum $k_{\vartheta \varsigma}^{2}$
\begin{eqnarray}
\label{row_54}
k_{\vartheta \varsigma}^{2} = \frac{{\cal M}_{\varsigma}^{2}}{d^{2}} +
k_{\vartheta}^{2} \; \; .
\end{eqnarray}
Finally, the quantity
\begin{eqnarray}
\label{row_55}
m_{4}^{2} = m^{2} + \frac{k_{\vartheta
\varsigma}^{2}}{c^2}
\end{eqnarray}
in Eq.(\ref{separation}) can be interpreted as the four-dimensional
square mass of a test particle in the flat
Min\-kow\-skian limit at infinity. In the case of the vanishing internal momentum $k_{\vartheta \varsigma} = 0$, the four and six-dimensional masses are equal $m_{4} = m$. If the six-dimensional mass is zero $m = 0$, then the
four-dimensional mass at spatial infinity would be solely of an internal origin: $m_{4} = \frac{|k_{\vartheta \varsigma}|}{c} $.

Now, let us consider the radial part $S_{r}(r)$, which can easily be read from Eq.(\ref{separation})
\begin{eqnarray}  \label{row_56}
\frac{d S_{r}}{d r} =  \left[ \frac{{\cal E}_{0}^2}{c^2} - \left( m^2 \, c^2 +
\frac{{\cal M}_{\Phi}^2}{r^2} \right) \, \frac{r}{r + A} -
k_{\vartheta \varsigma}^2 \, \left( \frac{r}{r + A} \right)^{\!\!
2} \right]^{\frac{1}{2}} \; , \nonumber
\end{eqnarray}
which after formal integration gives
\begin{eqnarray}  \label{row_57}
\!\!\!\!\! \!\!\!\!\!\!
S_{r} (r) = \int \! dr  \left[
\! \frac{{\cal E}_{0}^2}{c^2} - \left( \! m^2  c^2 + \frac{{\cal
M}_{\Phi}^2}{r^2} \right) \! \frac{r}{r+ A} -  k_{\vartheta
\varsigma}^2  \left( \! \frac{r}{r + A} \right)^{\!\! 2}
\right]^{\frac{1}{2}} \!  . \nonumber
\end{eqnarray}
The trajectory of a test particle is implicitly determined by the following equations
\begin{eqnarray}
\label{row_58}
\frac{\partial S}{\partial {\cal E}_{0}} = \alpha_{1} = - t + \frac{\partial S_{r} (r)}{\partial {\cal E}_{0}} \; ,
\end{eqnarray}
\begin{eqnarray}
\label{row_59}
\frac{\partial S}{\partial {\cal M}_{\Phi}} = \alpha_{2} = \Phi + \frac{\partial S_{r} (r)}{\partial {\cal M}_{\Phi}} \; ,
\end{eqnarray}
where $\alpha_{1}$ and $\alpha_{2}$ are constants and without loss
of generality, the initial conditions can be chosen so that
$\alpha_{1} = \alpha_{2} = 0$. In other words, integration of
Eq.(\ref{row_59}) gives the trajectory $r = r (\Phi)$ of a test
particle and Eq.(\ref{row_58}) provides the temporal dependence of
radial coordinate $r = r (t).$ These two relations determine the
trajectory $r = r (t)$ and $\Phi = \Phi (t)$ of a test particle
\cite{Rubinowicz-Krolikowski-1980}. \\
Implementing the above outlined procedure, we obtain (from
Eq.(\ref{row_58}) and  Eq.(\ref{row_57}))
\begin{eqnarray}  \label{row_60}
t = \frac{{\cal E}_{0}}{c^2}\,\int dr \left[ {\frac{{{{{\cal
E}_{0}}^2}}}{{ c^2}}} - {\left( {m^2}\,{c^2} + {\frac{{{{{\cal
M}_{\Phi}}^2}}}{{r^2}}} \right) \frac{r}{r + A}}
-  {k_{\vartheta \varsigma}^2} \, {{{\ \left( \frac{r}{r +
A} \right) }^{\! 2}}} \right]^{- \frac{1}{2}}
\end{eqnarray}
and consequently the radial velocity
\begin{eqnarray}
\label{row_61}
\frac{d r}{d t} =  \frac{c^2}{{\cal E}_{0}} \, \left[{\frac{ {{\
{{\cal E}_{0}}^2}}}{{c^2} }} - \left( {m^2}\,{c^2} +
{\frac{{{{{\cal M}_{\Phi}}^2}}}{{r^2} }} \right) \, \frac{r}{r +
A} -  {k_{\vartheta \varsigma}^2} \,
{{{\ \left(\frac{r}{r + A} \right) }^{\! 2}}}
\right]^{\frac{1}{2}} \; \; .
\end{eqnarray}
It is easy to note that the quantity
\begin{eqnarray}
\label{row_62}
\!\!\! m_{4}(r) = \sqrt{ m^2 \, \left( \frac{r}{r +
A} \right) + \frac{k_{\vartheta \varsigma}^2}{c^2} \, \left(
\frac{r}{r + A} \right)^2 }
\end{eqnarray}
can be interpreted as the mass in four-dimensional curved
space-time. The mass $m_{4}$ in Eq.(\ref{row_55}) is recovered as the
limit of $ m_{4}(r)$ for $r \rightarrow \infty$. Similar to our
previous discussion, if $k_{\vartheta \varsigma} = 0$ then
$m_{4}(r) = m \, \sqrt{\frac{r}{r + A}} $ and if
$m = 0$ then the mass $m_{4}(r)$ would have an internal origin
$m_{4}(r) =
\frac{|k_{\vartheta \varsigma}|}{c} \, \left( \frac{r}{r + A}
\right) $. It should also be emphasized that $m_{4}(r)$ does
depend on the ratio $r^{w} = \frac{r}{A} $ rather than separately
on $r$ and $A$, {\it which is a reflection of the scale-invariance of the model} (see also
Section~\ref{Scale invariance} and \cite{Dziekuje-za-neutron}).

Let us also notice that the non-negativity of the square of the
four-dimensional mass $m_{4}^{2}(r) \geq 0$ gives the first
condition on the value of $k_{\vartheta \varsigma}^{2}$
\begin{eqnarray}
\label{k2m24 minimum}
m_{4}^{2}(r) \geq 0 \; , \;\;\; {\rm so} \;\;\;\; k_{\vartheta \varsigma}^{2} \geq k_{\vartheta \varsigma m_{4}^{2}}^{2} \equiv - \, c^2 m^2 \, \frac{(r + A)}{r} \; .
\end{eqnarray}
If the entire
spacelike hypersurface is to be accessible for a particle, then from
the validity of the condition $m_{4}^{2}(r) \geq 0$,
the condition $k_{\vartheta \varsigma}^2~\geq~-m^{2}~c^{2}$ follows.
Otherwise, Eq.(\ref{k2m24 minimum}) is the condition for the
maximum value of the radius $r$
\begin{eqnarray}
\label{rm4}
r \leq r^{max} \equiv r^{max}(k_{\vartheta \varsigma m_{4}^{2} }^2)
\equiv - A/(1 + \frac{k_{\vartheta \varsigma m_{4}^{2} }^2}{m^2 \, c^2})\, .
\end{eqnarray}
{\it
It may appear strange that $k_{\vartheta \varsigma}^2 \geq - m^{2}
\,c^{2}$ by which some imaginary values of $k_{\vartheta
\varsigma}$ are also allowed. However,
one should not uncritically transfer
from four-dimensional intuitions (such like
$k_{\vartheta \varsigma}^2 \geq 0$ or $m^2 \geq 0$, although some
of them may turn out to be true) to e.g. the six-dimensional one but rather build on the safe ground
of known properties of four-dimensional sector i.e. $m_4^2 \geq 0$
in this case} \cite{Dziekuje-za-channel}. \\
\\
Now, using the Eq.(\ref{row_59}) and Eq.(\ref{row_57}) we obtain
\begin{eqnarray}  \label{row_63}
\!\!\!\!\!\!\!\!\!\!\!\!\!\!\!\!\!\!
\Phi = \int \left[{\frac{{{{{\cal E}_{0}}^2}}}{{c^2}}} - {\left(
{m^2}\,{c^2} + {\frac{{{{{\cal M}_{\Phi}}^2}}}{{r^2}}} \right)
\frac{r}{r + A}} -  k_{\vartheta
\varsigma}^2 {{{\ \left( \frac{r}{r + A} \right) }^{\!\! 2}}} \,
\right]^{- \frac{1}{2}} \, \frac{{\cal M}_{\Phi} \, dr}{r\,(r +
A)}
\end{eqnarray}
and consequently using Eqs.(\ref{row_63}) and (\ref{row_61}), the angular velocity of the particle reads
\begin{eqnarray}  \label{row_64}
{\Omega}_{t} = {\frac{d \Phi}{d t}} =  \frac{d r}{d t} \, {\frac{d \Phi}{d r}}
= {\frac{{{c^2}\,{\cal M}_{\Phi}}}{{{\cal E}_{0} \, r \,\left( r +  A  \right)
}}} \; \; .
\end{eqnarray}
Now, the proper angular velocity of the particle is equal to
\begin{eqnarray}  \label{row_65}
{\Omega}_{\tau} = {\frac{d \Phi}{d \tau}} = {\Omega}_{t} \, \sqrt{\frac{r + A}{r}} \; ,
\end{eqnarray}
where $\tau $ is the proper time and
$d \tau = \sqrt{g_{tt}} \,dt$ (where $g_{tt}$ is as in  Eq.(\ref{row_26})). \\
The transversal velocity of the particle (i.e. the component perpendicular
to the radial direction) is equal to (see Eq.(\ref{row_32}))
\begin{eqnarray}  \label{row_66}
v_{t} = {\Omega}_{t} \, r_{l}
\end{eqnarray}
and analogously, the transversal component of the proper velocity (written in
the units of the proper time $\tau$) is equal to
\begin{eqnarray}  \label{row_67}
v_{\tau} = {\Omega}_{\tau} \, r_{l} = v_{t} \, \sqrt{\frac{r + A}{r}}
\end{eqnarray}
where ${\Omega}_{t} $, ${\Omega}_{\tau} $ and $r_{l} $ are given
by Eqs.(\ref{row_64}),(\ref{row_65})~and~(\ref{row_32}),
respectively.\\
Let us rewrite Eq.(\ref{row_61}) in the following form
\begin{eqnarray}  \label{row_68}
\frac{dr}{dt} = \frac{c}{{\cal E}_{0}} \: \sqrt{{\cal E}_{0}^{2} -
{\cal U}_{eff}^{2} (r) }
\end{eqnarray}
where
\begin{eqnarray}  \label{row_69}
\!\!\!\!\!\!\!\!\!\!\!\!\!\!\!\!\!\!
{\cal U}_{eff} (r) = \left[ \left( m^{2} \, c^{4} + \left(
\frac{{\cal M}_{\Phi}}{r} \right)^{2} \, c^{2} \right) \,
\frac{r}{r + A} +  \left(
k_{\vartheta \varsigma}^{2} \,c^{2} \right) \left( \frac{r}{r + A}
\right)^{\!\! 2} \, \right]^{\frac{1}{2}} \!\! .
\end{eqnarray}
The function ${\cal U}_{eff} (r) $ plays the role of effective
potential energy in the meaning that the relation between ${\cal
E}_{0}$ and ${\cal U}_{eff} (r)$ determines the allowed regions of
the motion of the particle.\\
The proper radial velocity of the particle is equal to
\begin{eqnarray}
\label{row_70}
v_{r} = \frac{d r_{l}}{d \tau} = \frac{\sqrt{- g_{rr}} \, d r}{\sqrt{g_{tt}}
d t} = \frac{d r}{d t} \; ,
\end{eqnarray}
where in the last equality the relation $g_{tt} = - g_{rr} $ is used (see Eq.(\ref{row_31})).
So the radial velocity $dr/dt$ and the proper radial velocity $v_{r} = dr_{l}/d \tau $ are equal, and
one should stress that this property is fundamentally different from the analogous
relation for a black hole.

\subsection{Stable circular orbits.}

\label{Stable circular orbits}

Let us  consider for $A>0$  a stable circular orbit with
given values of ${\cal E}_{0}$, ${\cal M}_{\Phi}$ and
$k_{\vartheta \varsigma} $. The radius of this orbit can be
calculated from the following equations (see Eqs.(\ref{row_68}),
(\ref{row_69}),(\ref{row_70}))
\begin{eqnarray}  \label{row_74}
\frac{dr}{dt} = 0 \; ,
\end{eqnarray}
\begin{eqnarray}  \label{row_75}
\frac{dv_{r}}{dt}= \frac{d^{2} r}{dt^{2}} = \frac{dr}{dt} \frac{d}{dr}(\frac{dr}{dt}) = 0 \; .
\end{eqnarray}
From Eqs.(\ref{row_75}),(\ref{row_68}),(\ref{row_69}) we obtain
the implicit relation between the radius $r_{s}$ of the stable
circular orbit  (index $s$), the angular momentum of the particle ${\cal M}_{\Phi}$ and the internal ``total momentum'' $k_{\vartheta
\varsigma}$
\begin{eqnarray}
\label{row_76}
\!\! ({\cal M}_{\Phi})^{2} = \! \left(
m^2 \,c^2 + 2 \, k_{\vartheta \varsigma}^{2} ( \frac{r_{s}}{r_{s} + A} )
\right) \left( \frac{A}{2 \, r_{s} + A} \right) r_{s}^{2}
\end{eqnarray}
so
\begin{eqnarray}
\label{row_77}
{\cal M}_{\Phi} = \pm A^{1/2} \, r_{s}
\,\sqrt{\frac{(r_{s} + A) \, m^2 \, c^2 + 2 \, k_{\vartheta
\varsigma}^{2} \, r_{s} }{A^2 + 3 \,A \, r_{s} + 2 \,r_{s}^2}} \; \; .
\end{eqnarray}
It is easy to verify that because $k_{\vartheta
\varsigma} = const$, hence ${\cal M}_{\Phi} / r_{s} \rightarrow 0$ as $r_{s}$
tends to infinity. \\
Using Eqs.(\ref{row_74}),(\ref{row_68}),  (\ref{row_69}) we
calculate the total energy ${\cal E}_{0}$ of a particle moving along
the stable circular orbit of the radius $r_{s}$
\begin{eqnarray}
\label{row_78}
\!\!\!\!\!\!\!\!\!\!\!\!\!\!\!\!\!\!
\!\!\!\!\!\!\!\!\!\!\!\!\!\!\!\!\!\!
({\cal E}_{0})^{2}  =  \left[ m^2 \, c^4 +
\left(\! m^2 \,c^4 + 2 \, k_{\vartheta \varsigma}^{2} \, c^2 \,
\frac{r_{s}}{r_{s} + A}  \right) \frac{A}{2 \, r_{s} + A} \right]
\left( \frac{r_{s}}{r_{s} + A} \right)  + \, k_{\vartheta \varsigma}^{2}
c^{2} \! \left( \frac{r_{s}}{r_{s} + A} \right)^2
\end{eqnarray}
or in other words
\begin{eqnarray}
\label{row_79}
{\cal E}_{0} =  c^2 \sqrt{\frac{ r_{s}\, \left[ 2
\, (r_{s} + A)^{2} \,m^2 c^2 + 3 \, A \, r_{s} k_{\vartheta
\varsigma}^{2} + 2 \, r_{s}^2 \, k_{\vartheta \varsigma}^2 \right] } {c^2 \,
(r_{s} + A)^2 \, (2 \,r_{s} + A) } }  \, .
\end{eqnarray}
As the first example, let us consider the motion of a particle with
given values of
${\cal M}_{\Phi} \neq 0 $ and $k_{\vartheta \varsigma}$. Figure~1
illustrates the function ${\cal U}_{eff}(r) $ for different values
of ${\cal M}_{\Phi} $ and $k_{\vartheta \varsigma} = 0.$
\begin{figure}[pb]
\begin{center}
\subfigure{\includegraphics[angle=0,width=120mm]{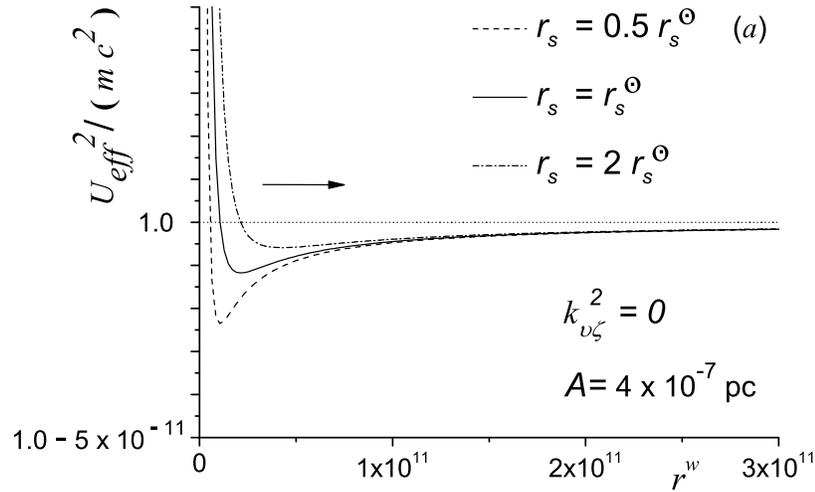}}
\end{center}
\vspace{-4mm}
\vspace*{8pt}
\caption{
The potential ${\cal
U}_{eff}$ (in units of $m c^{2}$) (see Eq.(\ref{row_69})) for
$k_{\vartheta \varsigma}^{2} = 0$ and different values of ${\cal
M}_{\Phi}$ as a function of the relative radius $r^{w} = r/A$.
Horizontal arrow above the curves, consistent with moving
of the minimum of ${\cal U}_{eff}$ to the right, denotes the direction of the increasing angular momentum ${\cal M}_{\Phi}$.  The minimum of the potential ${\cal U}_{eff}$ determines the radius $r_{s}^{w}$, and
hence the value of $r_{s} = r_{s}^{w} \, A$ of the stable circular orbit  (index $s$).
The radii $r_{s}^{w}$ of stable orbits corresponding to the minima of the
potentials depicted on Figure~1 are $1.0625 \; 10^{10}$, $2.125 \;
10^{10}$ and $4.25 \; 10^{10}$, respectively. The middle curve
(solid line) illustrates the effective potential for the stable circular orbit with a radius equal to the distance of the sun from the center of the dilatonic field $\varphi$ (with the parameter $A$ equal to $4. \, 10^{-7} \; {\rm pc}$) located at the center of the Milky Way, i.e. $r_{s}^{\odot} = 8.5 \; {\rm kpc}$.
 \label{f1}
}
\end{figure}
\begin{figure}[pb]
\begin{center}
\subfigure{\includegraphics[angle=0,width=120mm]{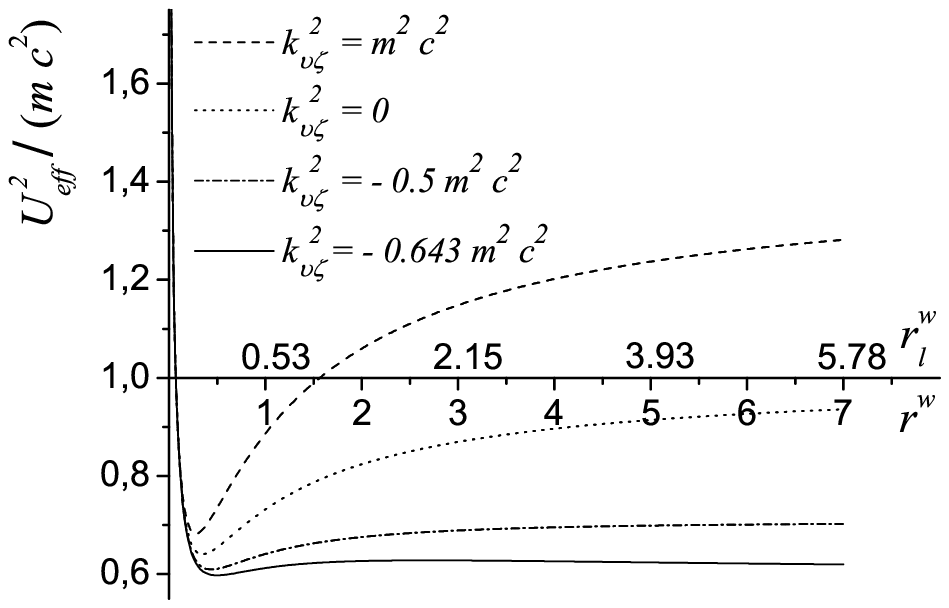}}
\end{center}
\vspace{-4mm}
\vspace*{8pt}
\caption{
The potential ${\cal U}_{eff}$ (in units of $m
c^{2}$) (see Eq.(\ref{row_69})) for different values of
$k_{\vartheta \varsigma}$  as a
function of the relative radius $r^{w} = r/A$.
For all curves, the value ${\cal M}_{\Phi}/A = 0.267 \, m \,c$  is chosen. This value of ${\cal M}_{\Phi}/A$ is the one
for the angular momentum of a particle on the stable orbit (the minimum of ${\cal U}_{eff}$) with
$ k_{\vartheta \varsigma}^{2} = - \,0.643 \,m^{2} c^{2}$
for which the radius
is equal to $r_{s}^{w} = 1/2$ (solid line). In this case
(solid line),
the maximum of ${\cal U}_{eff}$ is at the finite radius equal to
$r^{w} \approx 2.69$.
 \label{f2}
}
\end{figure}
Similarly, Figure~2 shows the function ${\cal U}_{eff} (r) $ for
different values of $k_{\vartheta \varsigma} $ with a fixed value of
${\cal M}_{\Phi}$.

For the particular system $r_{s}$ is  established and
$k_{\vartheta \varsigma}$ is the constant of motion.
Yet, there are some physically obvious conditions which in turn
restrict admissible values of the radii of stable orbits for a given
internal momentum $k_{\vartheta \varsigma}$. Namely, from Eq.(\ref{row_76}) we obtain
\begin{eqnarray}
\label{row_80}
\left( {\cal M}_{\Phi} \right)^{2} >
0 \; , \;\; {\rm so} \; \; \; k_{\vartheta \varsigma}^2 > k_{\vartheta \varsigma {\cal M}_{\Phi} }^2 \; ,
\end{eqnarray}
where
\begin{eqnarray}
\label{k min Mfi}
k_{\vartheta \varsigma {\cal M}_{\Phi} }^2 \equiv - \,
\frac{m^2 \,c^2}{2} \, ( \frac{r_{s} + A}{r_{s}} )
\end{eqnarray}
and from Eq.(\ref{row_78}) we obtain
\begin{eqnarray}  \label{row_81}
\left( {\cal E}_{0} \right)^{2} \geq
0 \;  , \;\;\; {\rm so} \;\;\;\; k_{\vartheta \varsigma}^2 \geq  k_{\vartheta \varsigma {\cal E}_{0}}^2 \; ,
\end{eqnarray}
where
\begin{eqnarray}  \label{kminE0}
k_{\vartheta \varsigma {\cal E}_{0}}^2 \equiv  - \,
m^2 \,c^2 \,\frac{2 \,(r_{s} + A)}{2 \,r_{s} + 3 \,A }
 \, ( \frac{r_{s} + A}{r_{s}} ) <  - m^2 \,c^2  \; .
\end{eqnarray}
It is easy to verify that condition (\ref{row_80})-(\ref{k min Mfi})
is stronger than (\ref{row_81})-(\ref{kminE0}) which in turn is the one that is stronger than (\ref{k2m24 minimum}), that is
\begin{eqnarray}  \label{trzy warunki na k}
k_{\vartheta \varsigma}^2 > k_{\vartheta \varsigma {\cal M}_{\Phi} }^2 > k_{\vartheta \varsigma {\cal E}_{0}}^2 > k_{\vartheta \varsigma m_{4}^{2}}^{2} \; ,
\end{eqnarray}
which means that the stability condition of the system is guaranteed by the first inequality in (\ref{trzy warunki na k}), i.e. by the condition (\ref{row_80})-(\ref{k min Mfi}).
Now, let us calculate the proper angular velocity and the proper transversal velocity of a particle moving along the stable circular orbit determined by
Eqs.(\ref{row_74}),(\ref{row_75}). Using Eqs.(\ref{row_64})-(\ref{row_67})
with ${\cal M}_{\Phi}$ and ${\cal E}_{0}$ given by Eq.(\ref{row_77}) and Eq.(\ref{row_79}), respectively, we obtain the proper angular velocity
\begin{eqnarray}
\label{row_82}
\Omega_{\tau}^{s} = {\frac{{{c^2}\,{\cal M}_{\Phi}}}{{{\cal E}_{0} \, r_{s} \,
\left( r_{s} + A  \right) }}}\, \sqrt{\frac{r_{s} + A}{r_{s}}}
\end{eqnarray}
and the proper tangent velocity of the particle
\begin{eqnarray}
\label{row_83}
v_{\tau}^{s} = {\frac{{{c^2}\,{\cal M}_{\Phi}}}{{{\cal E}_{0} \, \sqrt{r_{s} \,
\left( r_{s} + A  \right)} }}} \, \frac{r_{l s}}{r_{s}}  \; , \;
\end{eqnarray}
where $r_{l s}$ is given by Eq.(\ref{row_32})
for the stable
circular orbit. Looking at Eqs.(\ref{row_77}),(\ref{row_79}), it is not difficult to notice that $v_{\tau}^{s}$ again depends
on the ratio $r_{s}^{w} = \frac{r_{s}}{A} $ rather than on $r_{s}$
and $A$ independently.

From Eq.(\ref{row_83}) and Eqs.(\ref{row_77}),(\ref{row_79}) and for $m \neq 0$, one can notice that~if
\begin{eqnarray}
\label{condition on k2}
k_{\vartheta \varsigma}^{2} \propto {\cal C} \cdot m^2 \;\; , \;\;\;\; {\rm where} \;\;\; {\cal C} = constant
\end{eqnarray}
then $v_{\tau}^{s}$ does not depend explicitly on the mass $m$
of the particle.
However, unless the relation between the proportionality constant ${\cal C}$ and the square of the velocity of light $c$ is not unique,
from Eqs.(\ref{row_83}),
(\ref{row_77}) and (\ref{row_79}), it can be noticed that $v_{\tau}^{s}$ is
additionally parameterized by ${\cal C}$.
This is clearly the non-classical effect that influences the shape of the rotation curves (Figure~3). The only bounds on $\cal C$ that arise in the model follow from the constraints of causality \cite{Dziekuje-za-channel} (\ref{k2m24 minimum}), energy (\ref{row_81})-(\ref{kminE0}) and angular momentum  (\ref{row_80})-(\ref{k min Mfi}) analyzed above.
For example, the angular momentum  constraint (\ref{row_80})-(\ref{k min Mfi})
fulfilled on the entire spacelike hypersurface
gives, according to  Eq.(\ref{condition on k2}),
the condition ${\cal C} > {\cal C}_{bound}$ with the lowest bound ${\cal C}_{bound} = - c^{2}/2$
%
on which  $k_{\vartheta \varsigma {\cal M}_{\Phi} }^2/(c^{2} m^2) =  - 1/2 $,
so that the
internal momentum $k_{\vartheta \varsigma}$ fulfills the relation $k_{\vartheta \varsigma}^2/(c^{2} m^2)  >
k_{\vartheta \varsigma {\cal M}_{\Phi} }^2/(c^{2} m^2) = - 1/2$.
%
Although it is not excluded,  we do not yet know the rule that forces the proportionality constant ${\cal C}$ in Eq.(\ref{condition on k2})   to be the same for all test particles inside the configuration with the particular central dilatonic  field~$\varphi$. \\
Finally, if (phenomenologically) ${\cal C}$ in Eq.(\ref{condition on k2}) is  scaled with $r_{s}$, then
the profile of the proper tangent velocity $v_{\tau}^{s}$ given by Eq.(\ref{row_83}), and thus the rotation curves, could be modified. The systematic determination of such a rule fulfilled  by  ${\cal C}$ lies beyond the analysis of this paper.

%
%

\subsubsection{The case with $m \neq 0$.}

\label{case with diff zero}

From Eqs.(\ref{row_80})-(\ref{k min Mfi}) we notice that if
$k_{\vartheta \varsigma}^{2} > - \frac{1}{2}  m^2  c^2 $ then
all values of the radius $r_{s}$ (or $r_{l \, s}$, see Eq.(\ref{row_32}))
for stable orbits are allowed.
Figure~3 displays the rotation curves calculated according to
Eq.(\ref{row_83}) (in units of the velocity of light c) for
different values of $k_{\vartheta \varsigma}^{2}$.
It shows that the
whole effect is i) most pronounced in the region close to the center of the spherically symmetric configuration of the dilatonic field $\varphi$ and is, ii) except for the case of $k_{\vartheta \varsigma}^{2} < - \frac{1}{2} \, m^2 \, c^2 $
discussed below, extended from the center to the infinity.
For example, let us take the sun in the Milky Way.
Then, e.g. for $k_{\vartheta \varsigma} = 0$ and at the distance of $r_{s}^{\odot} = 8.5 \, {\rm kpc} $ from the center of the dilatonic field $\varphi$ that is overlapping with the center of the Milky Way, we obtain the contribution of $v_{\tau}^{s}$ (caused by the dilatonic field $\varphi$ with the parameter $A$ equal to $4. \, 10^{-7} \; {\rm pc}$)  to the total proper tangent velocity  equal to $ 1.45 \, {\rm km/s}$.
\begin{figure}[pb]
\begin{center}
\subfigure{\includegraphics[angle=0,width=120mm]{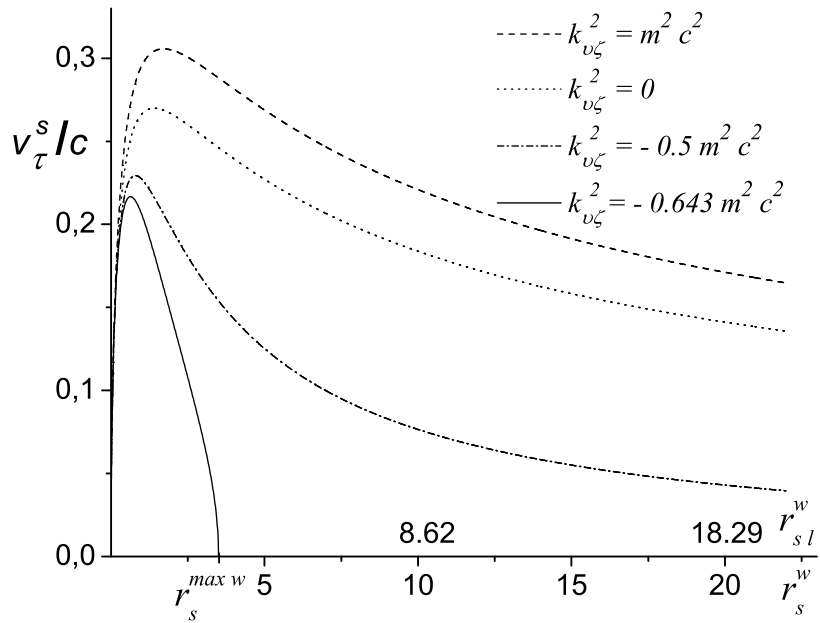}}
\end{center}
\vspace{-8pt}
\vspace*{8pt}
\caption{
The proper transversal velocity $v_{\tau}^{s}$ (see
Eq.(\ref{row_83})) of a test particle (in units of the velocity of
light c) moving along a stable circular orbit determined by
Eqs.(\ref{row_74}),(\ref{row_75}) as the function of
the relative radius $r_{s}^{w} = r_{s}/A$ for different (but fixed for
each curve) value of $k_{\vartheta \varsigma}^{2}$ and the
angular momentum ${\cal M}_{\Phi}$ calculated according
to Eq.(\ref{row_77}). If $k_{\vartheta \varsigma}^{2}
> - \frac{1}{2} \, m^2 \, c^2 $ then all values of the radius
$r_{s}^{w}$ (or $r^{w}_{s\, l} = r_{s\, l}/A $ cf. Eq.(\ref{row_32})) of stable orbits are allowed.
If $k_{\vartheta
\varsigma,min}^{2} <  k_{\vartheta \varsigma}^{2} <
- \frac{1}{2} \, m^2 \, c^2 $, then the stable orbits, for fixed
$k_{\vartheta \varsigma}^{2}$ and angular momentum ${\cal M}_{\Phi}$
chosen according to Eq.(\ref{row_77}), may exist only to the
radius $r_{s}^{max}$ (see Eqs.(\ref{r oraz rmax1})-(\ref{rmax1})). If $k_{\vartheta \varsigma}^{2} = - \,0.643 \,m^{2} c^{2}$  (as in
Figure~2 for the solid line) then $r^{max, w}_{s} \approx 3.5 \,$ and for $r^{w}_{s} \geq r^{max, w}_{s}$ there are no stable orbits.
For the limiting case $k_{\vartheta \varsigma}^{2} = - \frac{1}{2} \, m^2 \, c^2 \, $ we obtain $r_{s}^{max,w} \rightarrow \infty$.
 \label{f3}
}
\end{figure}
\\
If, however,
\begin{eqnarray}
\label{k przypadek niestabilny}
k_{\vartheta \varsigma,min}^{2} \equiv k_{\vartheta \varsigma {\cal M}_{\Phi} }^2 < k_{\vartheta \varsigma}^{2} < - \, \frac{1}{2} \, m^2 \, c^2
\end{eqnarray}
then according to Eqs.(\ref{row_80})-(\ref{k min Mfi}), the radius of the stable circular orbits has to fulfill the relation
\begin{eqnarray}
\label{r oraz rmax1}
r_{s} < r_{s}^{max} \; ,
\end{eqnarray}
where the limiting value is equal to
\begin{eqnarray}
\label{rmax1}
r_{s}^{max} \equiv r_{s}^{max}(k_{\vartheta \varsigma {\cal M}_{\Phi} }^2) \equiv - A/(1 + \frac{2 \, k_{\vartheta \varsigma {\cal M}_{\Phi} }^2}{m^2 \, c^2}) \, .
\end{eqnarray}
In other words, the condition (\ref{row_80})-(\ref{k min Mfi}) implies
that one cannot find any stable orbit with $r_{s} > r_{s}^{max}$
for a fixed $k_{\vartheta \varsigma}^{2}$,
which is from the range given by Eq.(\ref{k przypadek niestabilny})
and with the angular momentum ${\cal M}_{\Phi}$ chosen according to
Eq.(\ref{row_77}). In this case, the orbit with $r_{s} = r_{s}^{max}$ is the metastable one. For the limiting value, i.e. for $k_{\vartheta \varsigma}^{2} = - \frac{1}{2} \, m^2 \, c^2 $ we can notice that $r_{s}^{max} \rightarrow \infty$.  \\
\\
{\bf Note:} Above, we have seen that
(for $m \neq 0$) $k_{\vartheta \varsigma}^{2} = k_{\vartheta \varsigma {\cal M}_{\Phi} }^2 \equiv - \frac{1}{2} \, m^2 \, c^2 $  is the boundary between configurations of test particles having stable orbits on the entire spacelike hypersurface and configurations with unstable orbits beginning with the radius $r_{s}^{max}$ and upwards. Interestingly, in accordance with Eq.(\ref{row_54}), we can also notice that if the value $k_{\vartheta \varsigma}^{2} = - \frac{1}{2} \, m^2 \, c^2$ is not to appear in the result of the accidental cancellation,  then the square of the internal momentum $k_{\vartheta}$  and the internal angular momentum have to take one of the discrete possibilities, e.g.
%
\begin{eqnarray}
\label{kth and Mzet general}
\!\!\!\!\!\!\!\!\!\!\!\!\!\!\!\!
\left(\frac{k_{\vartheta}}{m \, c} \right)^{2} = - \jmath^2 +  \left(\frac{k_{\vartheta \varsigma}}{m \, c} \right)^{2}     \;\;\;\; {\rm and} \;\;\;\;  \frac{{\cal M}_{\varsigma}}{m \, c \, d} = \jmath \; ,
\;\;\;\; \jmath = 0,\frac{1}{2},1,\frac{3}{2},2,... \; ,
\end{eqnarray}
respectively, where  $\left(\frac{k_{\vartheta \varsigma}}{m \, c} \right)^{2} = -1/2 \, $ for the boundary stable configuration.
Yet, this choice  also agrees with the axial symmetry of the internal space (\ref{row_9}) around
the $\vartheta$ axis (compare also \cite{Dziekuje-za-neutron}). \\

\vspace{-2mm}

\subsubsection{A case with $m = 0 $ and $k_{\vartheta \varsigma}^{2}  \neq 0$.}

\label{case with zero}

In the case of $m = 0 $ and $k_{\vartheta \varsigma}^{2} \neq 0$, we can see from Eqs.(\ref{row_80})-(\ref{k min Mfi}),(\ref{row_76}) and Eqs.(\ref{row_81})-(\ref{kminE0}),(\ref{row_78}) that $k_{\vartheta \varsigma}^{2}
> 0$ and
all values of the radius $r_{s}$ of stable orbits are allowed.

\subsection{Radial trajectories, ${\cal M}_{\Phi} = 0$.}

\label{Radial trajectories}

In this case, we investigate the free motion of a test particle (with given
$k_{\vartheta \varsigma}^{2} $) along the geodesic $ \Phi = constant$   that
crosses the center of the gravitational field $g_{MN} $. Hence,  unless it is stated differently,  ${\cal M}_{\Phi} = 0$ in almost all of  Section~\ref{Radial trajectories}. From
Eqs.(\ref{row_70}) and (\ref{row_68}),(\ref{row_69}), we obtain
\begin{eqnarray}
\label{row_71}
\!\! v_{r}
= \frac{c}{{\cal E}_{0}} \, \sqrt{{\cal E}_{0}^{2} - \left(
{m^2}\,{c^4} \right) \, \frac{r}{r + A} - k_{\vartheta
\varsigma}^2 \, c^2 \, \left( \frac{r}{r + A} \right)^{\!\! 2} }
\, .
\end{eqnarray}
From this we may notice that for a particle which is initially
($t=t_{o}$) at rest ($v_{r}=v_{r_{o}}=0$) at the point $r = r_{o} \neq 0$, the
total energy is equal to (compare Eq.(\ref{row_62}))
\begin{eqnarray}  \label{row_72}
{\cal E}_{0} = \sqrt{ m^2 \,c^4 \, \left(
\frac{r_{o}}{r_{o} + A} \right) + k_{\vartheta \varsigma}^{2} \,
c^2 \, \left( \frac{r_{o}}{r_{o} + A} \right)^{\!\! 2} \, } = m_{4}(r_{o}) \, c^{2} \; .
\end{eqnarray}
Therefore, the particle is oscillating and crosses the center with velocity $
v_{r}(r=0) = c $ (at the center, the particle becomes massless $m_{4}(r=0)=0$).
From Eqs.(\ref{row_71}), (\ref{row_61}) and $d \tau = \sqrt{g_{tt}} \: dt$, we obtain that the acceleration of the particle is equal to
\begin{eqnarray}
\label{row_73}
\!\!\!\!
a_{r} = \frac{d v_{r}}{d \tau}
=  - A \,c^2 \,\frac{ m^2 \,c^4 + 2 \,k_{\vartheta \varsigma}^2
\, c^2 \,\left( \frac{r}{r + A} \right) } {2 \, {\cal E}_{0}^{2}
\, (r + A)^2} \sqrt{\frac{r + A}{r}} \, .
\end{eqnarray}

\subsubsection{A case with $m \neq 0$.}

\label{case with diif zero dwa}

From Eq.(\ref{row_73}) we see that the acceleration tends to minus
infinity at the center and when $k_{\vartheta \varsigma}^{2} \geq -
\frac{1}{2} \, m^2 \, c^2 $, it monotonously increases to the zero
value when $r$ is going to infinity. So, in this case, the
particle is attracted to the center for all values of $r$ (see
Figure~4).
\begin{figure}[pb]
\begin{center}
\subfigure{\includegraphics[angle=0,width=120mm]{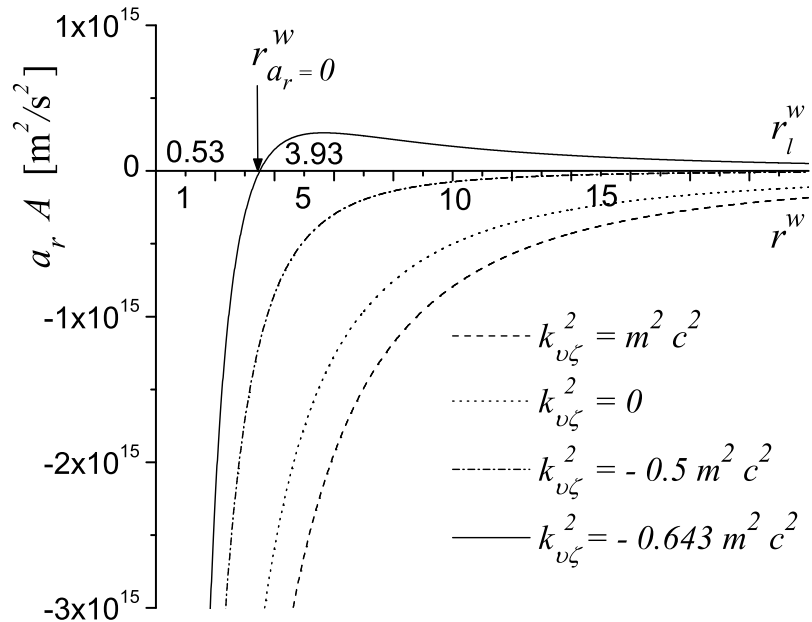}}
\end{center}
\vspace*{8pt}
\caption{
The acceleration
$a_{r}$ (in units of $1/A$) of a test particle moving along a radial
trajectory (see Eq.(\ref{row_73})). As in Figure~2, $k_{\vartheta \varsigma}^{2}$ for the solid line is chosen as equal to $ k_{\vartheta \varsigma}^{2} = - \,0.643 \,m^{2} c^{2}$. Consequently $a_{r} = 0$ for
the relative radius $r^{w} \equiv r/A = r_{a_{r} = 0}^{w}
\approx 3.5$ (see Eq.(\ref{ro dla a rownego 0})) and the ``particle'' is attracted
to the center for all values of $r^{w} < r_{a_{r} = 0}$ and is repelled
for $r^{w} > r_{a_{r} = 0}^{w}$ (see the text).
%
For other curves the
particle is attracted to the center ($a_{r} < 0$) for all values
of $r^{w}$.
\label{f4}
}
\end{figure}
\begin{figure}[pb]
\begin{center}
\subfigure{\includegraphics[angle=0,width=120mm]{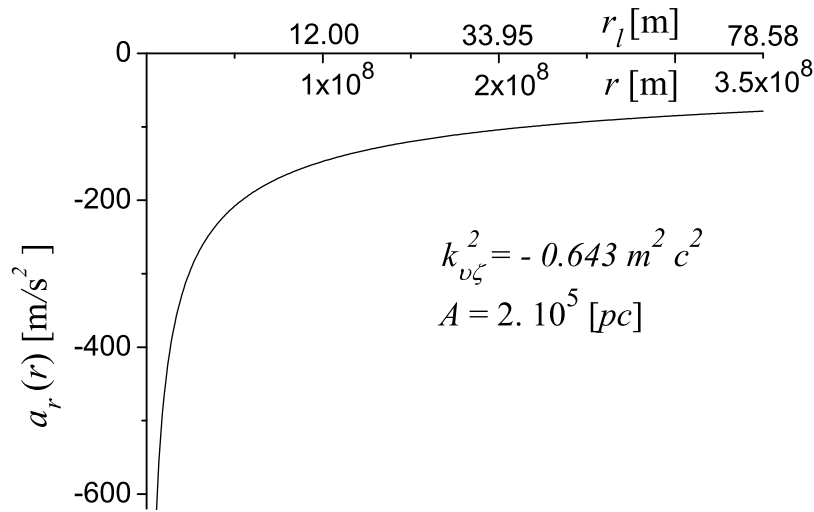}}
\end{center}
\vspace*{8pt}
\caption{
The radial acceleration $a_{r}$ of a test particle (see
Eq.(\ref{row_73})) for $ k_{\vartheta \varsigma}^{2} = - \,0.643 \,m^{2} c^{2}$ in the vicinity of the center of $\varphi$. The value $A = 2. \; 10^{5} \, {\rm pc}$ was chosen as typical for quasar-galaxy systems (see Table~1
in Section~\ref{prop}).
\label{f5}
}
\end{figure}\\
If $- m^{2} \, c^{2} \leq k_{\vartheta \varsigma}^{2}
< - \frac{1}{2} \, m^2 \, c^2 $ then from Eq.(\ref{row_73}) it follows that $r = r_{a_{r} = 0}$ with the finite value exists
\begin{eqnarray}
\label{ro dla a rownego 0}
r_{a_{r} = 0}  = - A/(1 + \frac{2 \, k_{\vartheta \varsigma  }^2}{m^2 \, c^2}) \,
\end{eqnarray}
for which $a_{r} = 0$  (compare Eq.(\ref{rmax1})). In this case, for
$r \leq r_{o} < r_{a_{r} = 0}$, the particle is attracted to the center with $a_{r} \rightarrow - \infty $ for $r \rightarrow 0 $ and for $r \geq
r_{o} > r_{a_{r} = 0}$, the particle is repelled from the center
and the acceleration $a_{r} \rightarrow 0$ when $r \rightarrow
\infty$ (see Figure~4).
In Figure~5 the radial acceleration
$a_{r}$ of the particle that is very close to the center of the
field $\varphi$ is presented. Because the solution is horizon free,
the achieved values of acceleration of the particle could
cause a visible point-like radiation (see
Section~\ref{noncosm-concl}).

\subsubsection{A case with $m = 0 $ and $k_{\vartheta \varsigma}^{2} \neq 0$.}

\label{case with zero dwa}

In this case, the requirement of ${\cal E}_{0}^{2} > 0$ implies $k_{\vartheta
\varsigma}^{2} > 0$ (see Eq.(\ref{row_72})). From Eqs.(\ref{row_71}),
(\ref{row_72})~and~(\ref{row_73}), we conclude
that the particle is attracted to the center ($a_{r} < 0$) for all values of
$r$.

\subsubsection{The causality condition.}

\label{The law of the lowest potential}

Let us suppose that we have $m_{4}^{2}(r) \geq 0$ on the entire spacelike hypersurface that is equivalent to the causality condition \cite{Dziekuje-za-channel}. Hence, from (\ref{k2m24 minimum})
we see that $k_{\vartheta \varsigma}^2 \geq$ $-m^{2}~c^{2}$ and from
the formula (\ref{row_69}), we obtain the condition for the effective potential at infinity, i.e. ${\cal U}_{eff}(\infty) \geq {\cal U}_{eff}(r=0) = 0$, with the equality if the limiting value $k_{\vartheta \varsigma}^2 = -m^{2}~c^{2}$ is chosen.
With $m_{4}^{2}(r) < 0$ for $r > r^{max}$, which according to (\ref{rm4}) is prohibited by the causality condition \cite{Dziekuje-za-channel} (\ref{k2m24 minimum}), the potential
${\cal U}_{eff}(r)$ becomes imaginary.
%
%
%

\subsubsection{A case with $m = 0 $ and $k_{\vartheta \varsigma}^{2} = 0$.}

\label{case with zero trzy}

Now in Eq.(\ref{row_71}) we have ${\cal E}_{0} = constant \neq 0$ and from Eqs.(\ref{row_71})~and~(\ref{row_73}), one can read  that $v_{r} = c$ and $a_{r}
= 0$ for all values of $r$. So, the particle which has both the
six-dimensional mass $m$ and the square of the total internal momentum
$k_{\vartheta \varsigma}^{2}$ equal to zero, which is a reasonable
representation of a photon for example, does not feel (except
changing the frequency according to Eq.(\ref{row_43})) the curvature of space-time when moving along the geodesic line crossing the
center. \\
On the other hand, for ${\cal M}_{\Phi} \neq 0 $, $m = 0 $
and $k_{\vartheta \varsigma}^{2} = 0$,  using Eq.(\ref{row_63})
and introducing formally the
parameter
$r_{m} = \frac{{\cal M}_{\Phi} \,c}{{\cal E}_{0}}$, we obtain the
trajectory of the particle
\begin{eqnarray}
\label{Phi-trajectory}
\!\!\!\! \Phi = \!\! \int \! \left[\frac{1}{r_{m}^{2}} -
\frac{1}{r^{2}} \, \frac{r}{r + A} \right]^{ - \frac{1}{2}}
\!\!\! \frac{dr}{r\,(r + A)} \;\;\; {\rm for} \;\;\;{\cal M}_{\Phi} \neq 0 \, .
\end{eqnarray}
When $A \rightarrow 0$ then the trajectory calculated according to the
above equation is a straight line $r = r_{m}/({\rm sin} \, \Phi)$
passing the center at the distance of $r_{m}$ (impact parameter).
On the other hand, light travelling in our space-time with $A\neq
0$ is deflected even in the absence of baryonic matter (see also
\ref{pericenter} on the Pericenter shift).\\
{\footnotesize
{\bf Remark}: {\it
\label{f-11}
We should use the eikonal equation
instead of the Hamilton-Jacobi equation for
$m = 0$ and $k_{\vartheta \varsigma}^{2} = 0$. However, a formal
(technical) substitution of $r_{m} = \frac{{\cal M}_{\Phi}
\,c}{{\cal E}_{0}}$ gives the same analytical
result~(\ref{Phi-trajectory}).}
}

\vspace{2mm}

\subsection{Redshift of radiation from stable circular orbits.}

\label{Redshift of the radiation}

\vspace{2mm}

Let us suppose for simplicity's sake that the observer is located far
away from the center of the system at a distance far bigger than
the size of the system, so its peculiar motion with respect to the
center of the system is negligibly small (see Figure~6).
Let us also assume that the dynamical time scale $t_{dyn}$ is greater than the characteristic timescale $t_{obs}$ over which the observations are performed i.e. $t_{dyn} > t_{obs}.$ In such a case, the observed
motion of the ``luminous particle'' is seen only as instantaneous redshift or blueshift. If the motion of the ``particle'', which is the  source, takes place along the stable circular orbit, then the kinematical Doppler shift is equal to (see Figures~6,7)
\begin{eqnarray}
\label{row_z1}
z_{D} = \sqrt{\frac{c - v_{ \tau}^{s} sin \Phi ( \tau ) } {c +
v_{\tau}^{s} sin \Phi ( \tau )} } - 1 \; ,
\end{eqnarray}
where the angle $\Phi ( \tau ) = \int_{0}^{\tau} \, \Omega_{ \tau}^{s} \,d \tau $ (see Eq.(\ref{row_82}))
is coun\-ted from the direction to observer, i.e.
$\Phi(\tau=0) = 0$, as in Figure~6, and the proper tangent velocity $v_{\tau}^{s}$ is given by Eq.(\ref{row_83}).
\\
Now, the gravitational redshift according to Eq.(\ref{row_46}) is equal to (see Figure~7)
\begin{eqnarray}
\label{row_z2}
z_{g} = \frac{\lambda_{obs}}{\lambda_{\sigma}} - 1 =
\frac{\omega_{\sigma}}{ \omega_{obs}} - 1 = \sqrt{\frac{r_{s} +
A}{r_{s}}} - 1 \;  ,
\end{eqnarray}
where $r_{\sigma}=r_{s}$.
\begin{figure}[pb]
\begin{center}
\vspace{-5mm}
\subfigure{\includegraphics[angle=0,width=80mm]{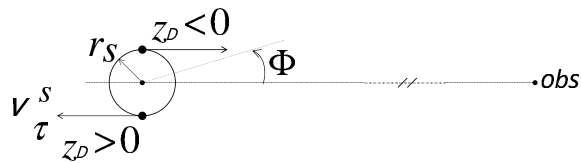}}
\end{center}
\vspace*{8pt}
\vspace{-15mm}
\caption{The Doppler shift: A source moving along the stable orbit of radius $r_{s}$. When it is moving
away form the observer (which is at infinity), then the instantaneous angle is equal to $\Phi=3/2 \, \pi$ and
$z_{D} > 0$, i.e. the source is (maximally) kinematically redshifted for the observer. When the source is moving
towards the observer, the instantaneous angle is equal to $\Phi=\pi/2$ and $z_{D} < 0$, i.e. the source is (maximally) kinematically blushifted.
\label{f6}
}
\end{figure}\\
It is not difficult to see that the combined effect of these
redshifts is as follows (see Figure~8)
\begin{eqnarray}
\label{row_z3}
z = \left( z_{g} + 1 \right) \, \left( z_{D} + 1 \right) - 1 \, \, .
\end{eqnarray}
For a source moving along the stable orbit
away form the observer (which is at infinity), the instantaneous angle is equal to $\Phi=3/2 \, \pi$ (see Figure~6) and from Eq.(\ref{row_z1}), we obtain that $z_{D} > 0$, i.e. the source is kinematically redshifted and the combined effect given by Eq.(\ref{row_z3})  obviously has a positive value of the redshift~$z$ (see the left-hand side of Figure~8). \\
In the case of the source moving
towards the observer, the instantaneous angle is equal to $\Phi=\pi/2$ (see Figure~6) and $z_{D} < 0$, i.e. the source is kinematically blushifted. Yet, we notice (see the right-hand side of Figure~8) that for all values of $k_{\vartheta \varsigma}^{2}$, but small enough values of $r_{s}$,  even in this case
the combined effect given by Eq.(\ref{row_z3}) results in a positive value of the redshift $z$. Moreover, for sufficiently small values of
$ k_{\vartheta \varsigma}^{2} < - \frac{1}{2} \, m^2 \, c^2 $, e.g. for  $ k_{\vartheta \varsigma}^{2} = - \, 0.643  \, m^2 \, c^2$ as in Figure~8,  even the whole  effect of the Doppler blueshift can be hidden behind the gravitational redshift. Because  there are no stable orbits for
$k_{\vartheta \varsigma}^{2} = - \, 0.643  \, m^2 \, c^2 $ for $r^{w}_{s} > r^{max, w}_{s}  \approx 3.5 \,$  (see Eq.(\ref{rmax1})), hence for the metastable point $r^{max, w}_{s}$
the breaks of the curves (with solid lines) plotted on both the right- and left-hand side of Figure~8 occur. \\
To summarize, for the square of the total internal
momentum $ k_{\vartheta \varsigma}^{2}$ that fulfills relation (\ref{k przypadek niestabilny}), we obtain
in accordance with  Eqs.(\ref{r oraz rmax1})-(\ref{rmax1}) that
for all allowed stable radii of the moving source, the possible Doppler blueshift is hidden behind the gravitational redshift caused by the dilatonic central field (see  Figure~8).
\begin{figure}[top]
\vspace{-2mm}
\begin{center}
\subfigure{\includegraphics[angle=0,width=120mm]{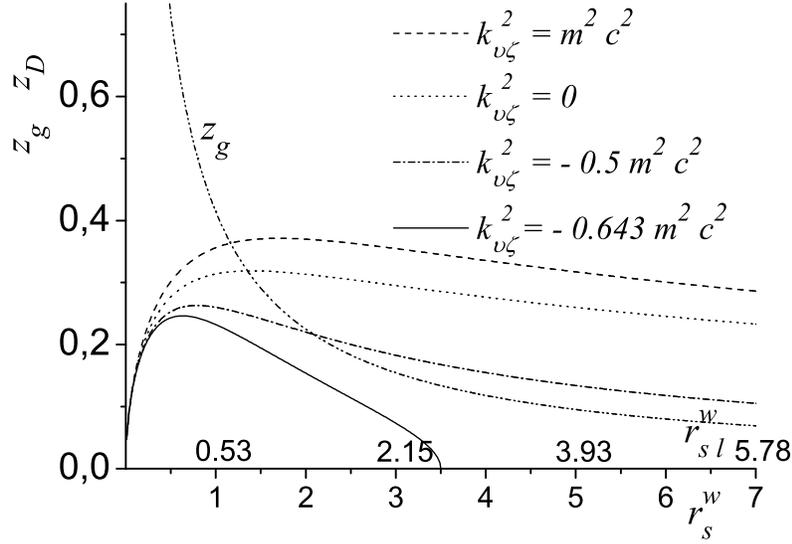}}
\end{center}
\vspace*{8pt}
\vspace{-2mm}
\caption{
The (maximal) Doppler shift
$z_{D}$ caused by motion of the particle along a stable
circular orbit (see Eq.(\ref{row_z1})) for different values of
$k_{\vartheta \varsigma}^{2}$.
The curve $z_{g}$ denotes the gravitational redshift (see
Eq.(\ref{row_z2})).
\label{f7}
}
\end{figure}
\begin{figure}[pb]
\vspace{-2mm}
\begin{center}
\subfigure{\includegraphics[angle=0,width=120mm]{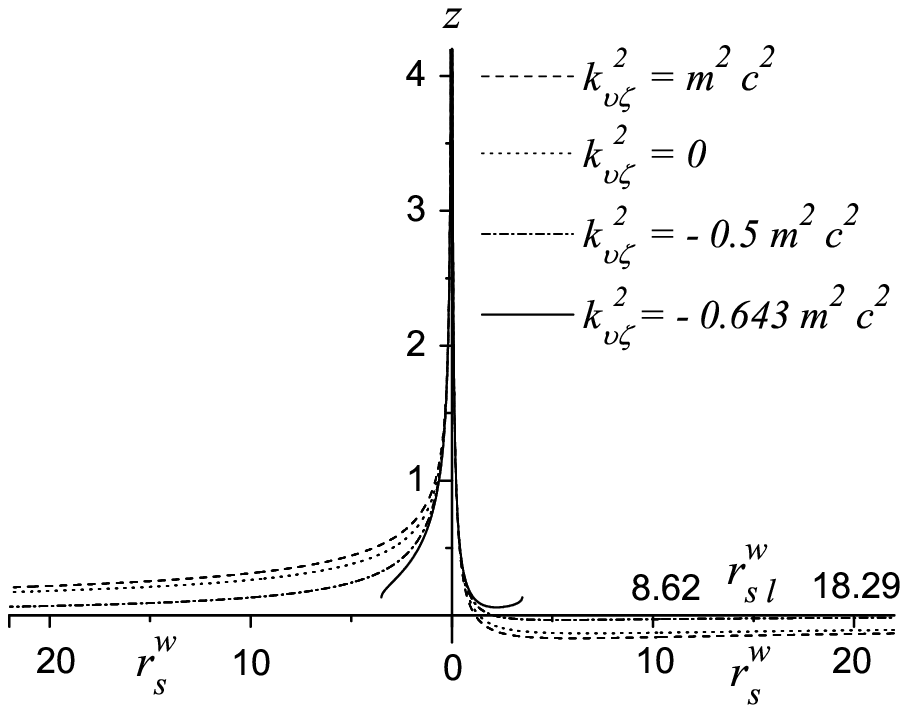}}
\end{center}
\vspace*{8pt}
\vspace{-2mm}
\caption{
The combined effect of the gravitational and
Doppler redshifts (see Eq.(\ref{row_z3})). For simplicity's sake we consider the limiting case when the observer is
situated at infinity. Thus, the curves on the right-hand side of the picture correspond to the sources for which $\Phi=\pi/2$, moving
towards the observer, whereas the curves on the left-hand side correspond to the sources for which $\Phi=3/2\, \pi$, running
away from the observer.
\newline
%
For small enough values of $r_{s}$, the combined Doppler and gravitational effect (\ref{row_z3}) has a positive value of the redshift $z$ for all values of $k_{\vartheta \varsigma}^{2}$. Yet, for sufficiently small values of $ k_{\vartheta \varsigma}^{2} < - \frac{1}{2} \, m^2 \, c^2 $, e.g. for  $ k_{\vartheta \varsigma}^{2} = - \, 0.643  \, m^2 \, c^2$,  even the whole
Doppler blueshift can be hidden behind the gravitational redshift. The breaks of the curves (with solid lines) for
$ k_{\vartheta \varsigma}^{2} = - \, 0.643  \, m^2 \, c^2 $ are at $r^{max, w}_{s} \approx 3.5 \,$, as for $r^{w}_{s} > r^{max, w}_{s}$ there are no stable orbits (see Eqs.(\ref{r oraz rmax1})-(\ref{rmax1})).
\label{f8}
}
\end{figure}

For example, let us choose $A = 96 \, {\rm pc}$ as the characteristic value for a system of pairs of galaxies (see 5th row in Table~1).
When one of the galaxies in the pair has a  four-dimensional mass equal to e.g. $m_{4} \approx 3.53 \times 10^{11} \, M_{\odot} $ and a radius of its stable orbit equal to $r_{s} = 2000 \, A = 192 \, {\rm kpc}$ and
$k_{\vartheta \varsigma}^{2} = - 0.500235  \, m^2 \, c^2$, where $m \approx 5 \times 10^{11} \, M_{\odot}$ is the six-dimensional mass of the galaxy, then the value of its total proper tangent velocity that is equal to $v_{\tau}^{s} \approx 36.6$ km/s
is obtained. The gravitational redshift (\ref{row_z2}) of the galaxy is equal to $z_{g} \approx 0.00025$
and the maximal value of the Doppler redshift (\ref{row_z1}) is equal to  $z_{D} \approx 0.00012$.
Then, the combined effect of the gravitational and Doppler redshifts (see Eq.(\ref{row_z3})) is equal to $z \approx 0.00037$ and $z \approx 0.00013$ for a galaxy moving along the stable orbit
away from and towards the observer, respectively.
Let us notice that, besides the huge value of (the auxiliary mass) $M$ connected with the value of the parameter $A$ of the gravito-dilatonic configuration  (see 5th row in Table~1),
the four-dimensional mass $m_{4}$ of the galaxy forms approximately 70.7\% of its six-dimensional mass, which is responsible for the kinematics of the galaxy in its  motion around the center of the background gravito-dilatonic field  configuration. The remaining 29.3\% mass is
connected with the galaxy total internal momentum $k_{\vartheta \varsigma}$. Thus, in some cases the mass $m$, which enters into the six-dimensional equations of motion could be bigger than the four-dimensional mass $m_{4}$, which could be perceived in the four-dimensional space-time.

\section{Scale invariance}

\label{Scale invariance}

Let us rewrite the Hamilton-Jacobi equation (see
Eq.(\ref{row_50})) in the following way
\begin{eqnarray}
\label{row_si1}
\!\!\!\!\!\!\!\!\!\!\!\!\!\!\!
\!\!\!\!\!\!\!\!\!\!\!\!\!\!\!
& &  \frac{r^{w} + 1}{r^{w}} \left( \frac{\partial
S}{\partial t^{w}} \right)^{\! 2} - \frac{r^{w} + 1}{r^{w}} \:
\left( \frac{\partial S}{\partial r^{w}} \right)^{\! 2} -
\frac{1}{(r^{w})^{2}} \: \left( \frac{\partial S}{
\partial \Theta} \right)^{\! 2} -  \frac{1}{(r^{w})^{2} \: sin^{2} \Theta} \: \left(
\frac{\partial S}{\partial \Phi} \right)^{\! 2} \nonumber \\
\!\!\!\!\!\!\!\!\!\!\!\!\!\!\!
\!\!\!\!\!\!\!\!\!\!\!\!\!\!\!
&-&
\frac{1}{(d^{w})^{2}} \: \frac{r^{w}}{r^{w} + 1} \:
\frac{1}{cos^{2} \vartheta} \: \left( \frac{\partial S}{\partial
\vartheta } \right)^{\! 2} -  \frac{1}{(d^{w})^{2}} \: \frac{r^{w}}{r^{w} + 1} \: \left(
\frac{
\partial S}{\partial \varsigma} \right)^{\! 2} - ( l^{w} \, c )^{2} = 0 \; ,
\end{eqnarray}
where
\begin{eqnarray}  \label{row_si2}
r^{w} = \frac{r}{A} \; , \; \; t^{w} =
\frac{ c\;t }{A} \; , \; \; d^{w} = \frac{d}{A} \; , \; \; l^{w} =
m \, A \; .
\end{eqnarray}
In fact Eq.(\ref{row_si1}) is scale invariant under
\begin{eqnarray}
\label{A invariance}
A \rightarrow \alpha \; A \; , \;\;
r \rightarrow \alpha \; r \; , \;\; d \rightarrow \alpha \; d  \; , \;\;  c\, t \rightarrow \alpha \; c\, t \;
\end{eqnarray}
only if this transformation is supplemented by
\begin{eqnarray}
\label{A transformation for mass}
m \rightarrow \frac{1}{\alpha} \, m \; ,
\end{eqnarray}
where $\alpha$ is the parameter of the transformation.
Systems with different values of $r$, $A$, $d$, $m$ and $t$ have the same physical properties provided the values of $r^{w}$, $t^{w}$, $d^{w}$ and $l^{w}$ are the same.
This means that whenever the dilatonic field $\varphi$ is present, and unless the symmetry (\ref{row_dzialanie-EH-fi})-(\ref{row_2}) is broken, the classical picture of the world follows the same patterns from the micro- to the marco-scale only if the complete integral of the system $S(t^{w},r^{w};\,d^{w},l^{w})$ does not change. \\
Transformation (\ref{A invariance}) is the symmetry of the action (\ref{row_dzialanie-EH-fi})-(\ref{row_2}).
%
%
%
This is the invariance of the coupled Einstein
(\ref{row_3}) and Klein-Gordon (\ref{row_5}) equations, but it
is not the invariance of their solution given by
Eqs.(\ref{varphi solution}) and (\ref{row_26}).
This conclusion can also be drawn from Eq.(\ref{row_24}).
Hence, we notice that the massless dilatonic
field $\varphi$ is the Goldstone field \cite{GSW}. \\
%
%
%
It would seem that invariance (\ref{A invariance}) inevitably leads
to the invariance given by Eqs.(\ref{A invariance})-(\ref{A transformation for mass})
of the Hamilton-Jacobi equation (\ref{row_si1}).
%
%
Yet, the inclusion of any term with mass $m \neq 0$ into the model given
by Eq.(\ref{row_dzialanie-EH-fi}) will inevitably lead to the
gravitational coupling of this mass
with the metric tensor
part of the Lagrangian, causing the breaking of the original invariance (\ref{A invariance}).
Hence, the conclusion emerges that there appears a relation between
the four-dimen\-sional mass $m_{4}$ of the particle and the parameter $A$ \cite{Dziekuje-za-neutron}. This
results from the Einstein equations which fix the value of $A$
after the self-consistent inclusion of the particle
%
%
with mass $m$ into the configuration of  fields. This means that the symmetry connected with the scaling (\ref{A invariance}) of $A$ is broken.
%
%
%


\vspace{-1mm}

\section{Conclusions and perspectives}

\label{noncosm-concl}

\vspace{-1mm}

In this paper the non-trivial, one parametric
six-dimen\-sional ``background'' solution given by Eqs.(\ref{row_26}) and
(\ref{varphi solution}) with the spherical symmetry in the
Minkowski directions of coupled Einstein
and Klein-Gordon equations has been presented   \cite{LORD-Biesiada-Syska-Manka,Dziekuje-Ci-Panie-Jezu-Chryste}. It consists in adding to the six dimensional gravi\-ty a mass\-less dilatonic field, whose variation in the radial
direction compensates self-consistently for the curvature. As a self-consistent \cite{Dziekuje-za-self-consistent-state} and
non-iterative one,
the solution is also non-pertur\-ba\-tive. Here,
the dilatonic ``basic'' field $\varphi$ forms a kind of the ground field and the gravitational field (metric tensor) $g_{M N}$ is the self-field \cite{Dziekuje-Ci-Panie-Jezu-Chryste,JacekManka,Dziekuje-Jacek-nova-2,Dziekuje-za-self-consistent-state}. 
The gravitational component in the
four-dimen\-sio\-nal space-time is asymptotically flat but
fundamentally different from the Schwarz\-schild solution.
Locally, the space-time has the topology of the Minkowki (1,3) $\times$ 2-dimen\-sio\-nal internal space of the varying size,
i.e. the six-dimensional world is compactified in a
non-homo\-geneous manner.
Next,  the motion of a test particle in the six-dimensional space-time of the
obtained gravito-dilatonic Kaluza-Klein ty\-pe model
was examined.
Additionally, in Section~\ref{case with diff zero} (see Eq.(\ref{kth and Mzet general})) it was pointed out that on the stability boundary, the relative internal angular momentum of the particle takes, beginning with zero, all consecutive values with the one half step, i.e.
$\frac{{\cal M}_{\varsigma}}{m \, c \, d} = \jmath$, where  $\jmath = 0,\frac{1}{2},1,\frac{3}{2},2,...\; $.

In Section~\ref{stability of the background} both the problem of the stability of the background
solution and the spectrum of the Kaluza-Klein type excitations were  discussed but additional
ana\-lysis of the problem can be
found in \cite{Dziekuje-za-neutron},
where the Fisher-statistical reason for six-dimen\-sio\-nality of the space-time was also discussed.
Nevertheless, since to date we do not see any well-established experimental evidence of the multi-dimen\-sionality of the world \cite{standard-extra-dimensions-in-LHC}, hence,  from the point of
view of practical purposes, the four-dimensional consistency
alone, which is possessed by a multi-dimensio\-nal model, cannot be perceived
as a very important reason for its acceptance. \\
Therefore, the main physical applicability of the background solution (\ref{row_26}),(\ref{varphi solution}) presented in Section~\ref{prop}  
refers to its astrophysical and
cosmological signatures for some relatively
tight systems.
Thus, the motion of a test particle in this
background configuration was analyzed. The background solution
is parameterized by the parameter $A$, which (especially for $r >> A$)  has similar dynamical consequences as the mass $M = A \, c^2/(2 \, G)$ (see Note below Eq.(\ref{row_32})).
The significance of higher terms in the metric tensor $g_{MN}$ expansion can be different than in e.g. the Schwarzschild solution case (see e.g.
\ref{pericenter} on the pericenter shift).
Hence, the existence of the self-consistent gravito-dilatonic configuration
(\ref{varphi solution}),(\ref{row_26}),
which is perceived by an observer in the same way as  invisible
matter, requires taking into account
the nonstandard defi\-nition of the system
as  one which is both under the self-consistent (Eqs.(\ref{row_5}),(\ref{row_3})) and non-self-consistent (Eq.(\ref{row_49})) influence of the gravitational interaction. It was performed
both on the astrophysical (see Section~\ref{motion}) and on the  physics of one particle \cite{Dziekuje-za-neutron} levels.
%
%

As the central object does not
possess a horizon (as the  black-hole does), which is the
frequent characteristic of more than four-dimensional solutions \cite{Wesson},
thus one of
the conclusions from the analysis of this paper is that the visible total redshift-blueshift asymmetry of the radiation from e.g. flows of sources of matter  (which consist of test particles)
is possible only if they
are located sufficiently close to the center of the dilatonic field $\varphi$ (Section~\ref{Redshift of the radiation}).
Therefore, the radiation from rotating matter, e.g. around a galaxy center, can originate from an area smaller than that connected with the corresponding Schwarzschild radius  $(2 \,G/c^2) M$ (see Note below Eq.(\ref{row_32})) of a black hole.
%
%
%
%
%
%
%
%
Yet, the observations of predominantly redshifted flows in the presented model can be (see Section~\ref{motion}) explained with  special effectiveness for the motion along circular orbits with the internal
momentum squared, which fulfills the relation
$k_{\vartheta \varsigma}^2 < - \, m^{2} \, c^{2} /2$. That is, in this case in the model even the whole Doppler effect connected with the blue wing moving towards us can be hidden below the gravitational redshift effect (see Figures~7,8). Thus, real flows of the matter of some objects can be directed towards us. In fact,
all galaxy- and quasar-like objects have to possess
proper $k_{\vartheta \varsigma}^2$ and four-dimensional square masses (\ref{row_55})
in order for them to be observed as redshifted only.

Looking from the above-mentioned perspectives, two classes of phenomena are invoked below. \\
%
%
%
%
Firstly,
the statistical analysis performed for the X-ray spectroscopy  points to a need for the relativistic effects to be included when modeling the emission from the rotating matter
around the galactic centers \cite{Brenneman-Reynolds}.
The analysis for the nucleus of the Seyfert
galaxy MCG-06-30-15  \cite{Brenneman-Reynolds-1}  was followed by the one for the nuclei of the galaxies MCG-6-30-15, NGC 4593,  3C 120, MCG–6-30-15, MCG–5-23-16, NGC 2992, NGC 4051, NGC 3783, Fairall 9, Ark 120 and 3C 273   \cite{Brenneman-Reynolds}. The conclusion was that the relativistic effects, such as the central object spin and the relativistic velocities of flows,
seem to become important in shaping the overall
emission line profile from the inner part of
the rotating matter of accretion disks,
i.e.
from
radii which are probably smaller than $\sim 20 \, A$.
%
%
%
%
The characteristics of the observed
emission lines can
correspond to a velocity of
$\sim 10^{5} \,$ km/s which, in agreement with the analysis in  Section~\ref{Stable circular orbits}, corresponds to a few values of $A$  (see Figure~3).
The observed
lines are asymmetric,
%
i.e.  strong
redshifted emission lines were detected and weaker (or none) blue-shifted ones were seen \cite{Tanaka-1995}. \\
%
%
%
%
%
%
%
Secondly, there are also other phenomena which call for explanations and are hard to understand from the point of view of the standard lore. One class of such problems is connected with the nature of the redshift of galaxies and quasars. It has been known for a long time that the visible associations of quasars and galaxies have been observed where the components have widely discrepant redshifts \cite{Arp-Giraud-Sulentic-Vigier-1983}.
We can think of a simple
explanation for such seemingly strange phenomena in terms of the
presented model.
That is, let us imagine that a quasar--galaxy system is captured by the 
gravito-dilatonic configuration of the  field $\varphi$ (see 6th row in Table~1).
Supposing that both the quasar and
galaxy are moving along  stable orbits,
we
apply the results obtained in Section~\ref{motion}. If the things are so arranged that the quasar (or a group of them) is closer to the center of the background gravito-dilatonic field configuration,  then it has a redshift greater than the galaxy located peripherally.
As  has been illustrated in Figures~7,8, the smaller the radius $r_{s}^{w}$ of the stable orbit of the quasar is the more significant this effect also is. \\
\\
{\footnotesize
{\bf Remark}: {\it
Nevertheless, quasar-like objects are also seen in the vicinity of the central parts of some galaxies. This could be a sign that a galaxy nucleus consists of the background gravito-dilatonic field configuration  (see 3rd row in Table~1).
The visibility of bigger values of the redshift of the luminous matter from the inner part of this galaxy nucleus could be blocked by the clumping of the matter orbiting and infalling on it.}
}
\\
\\
Recently a large quasar group (LQG) with 73 members with a huge redshift range $1.1742 \leq z \leq 1.3713$ was identified in the DR7QSO catalogue of the Sloan Digital Sky Survey \cite{large-quasar-group}. Such a spread of quasars' redshift can be
understood as  resulting from bounding
them in one node of the gravito-dilatonic configuration.\\
{\bf Note:} The four-dimensional
part of the metric tensor (\ref{row_7}) possesses a
spherical symmetry. Thus, it should be emphasized that in  reality the orbits
do not necessarily lay on one plane. Next,
the instantaneous position of an object in space is observed as projected onto a two-dimensional sphere along the light's geodesics in the curved space-time.
This sphere is then seen as a plane perpendicular to us only. This can obviously place some objects with bigger $r_{s}$ seemingly closer to the center of the background gravito-dilatonic field configuration. Finally, if constant ${\cal C}$ in Eq.(\ref{condition on k2}) does not have the same value for all orbiting objects then their proper tangent velocities, (\ref{row_83}), depend on ${\cal C}$
and  the observational
situation gets more complex.
\\
%
%
%
When the distance from the center is a fraction of $A$,
the combined gravitational and Doppler redshift given by Eq.(\ref{row_z3}) becomes the sloping function of $r_{s}^{w}$  thus disclosing  the existence of a big redshift discrepancy.
Such a possibility also has an attractive feature that if the quasar--galaxy constitutes a kind of
binary system, then the visible bridge of matter connecting the galaxy and quasar  may be explained as a sign of an infall (from the galaxy to the center of the field $\varphi$) of matter that is passing through the quasar. \\
Let us also recall that near the center of the field $\varphi$ the space-time curvature (\ref{row_24}) is large, strongly accelerating  (see Eq.(\ref{row_73})) the infalling particles. Therefore, even if the  diameter of the quasar size is of $\sim 100$  AU only then due to the fact that the quasar is located near the center of $\varphi$, does the existence of
the highly nonuniform and anisotropic gravitational field across the quasar interior cause the appearance of large tidal forces inside it. This sheds some light on the nature of the quasar emission as connected with the ultrahigh relative acceleration of particles in its interior, also charged ones, which are mainly electrons. The idea briefly outlined above deserves further, deeper considerations. \\
Meanwhile,
if a
galaxy was close to the center of the field $\varphi$, let's say in a stable orbit with $r_{s} \sim \,A/2$ (see Figure~4,5), then due to its size it would have been torn into pieces by the tidal forces of the gravito-dilatonic center. This means that the only possibility for a ``galaxy" to possess a bigger redshift is to be placed at the center of the background gravito-dilatonic field configuration with the central, highly redshifted opaque inner part of this ``galaxy" being
severely overshadowed by the more peripheral luminous matter. This  overshadowing
could account for its
faintness.
However, this means that e.g. an object like UDFj-39546284 \cite{Bouwens} is not really a ``usual" galaxy. Nevertheless, the real obstacle in the precise identification of these objects is the lack of their observed detailed structure.
With
the lack of the confirmation that objects like these emit all wavelengths both shorter than 1.34 ${\rm \mu m}$  and longer than  1.6 ${\rm \mu m}$,  the spectroscopy analysis  \cite{Bouwens,Graham} increases
the doubt about their true galaxy structure .
%
%

What is more, there is a great deal of evidence that
redshift is at least partially the intrinsic property of
galaxies and quasars that is apparently quantized
\cite{Arp-1986,Tifft-1988,Tifft-93,Arp-2013,Hansen-2014}.
Its nature is not yet understood
and an analysis of this phenomenon is not covered by the presented, classical mo\-del.
The solution to the problem may require
e.g. the modification of the background metric so that the modified  one possesses additional minima in the component $g_{rr}$.
This could be achieved by coupling the Einstein equations (\ref{row_3}) to the one which replaces the Klein-Gordon equation (\ref{row_5}) and only then solving the Hamilton-Jacobi equation used in this paper for the description of the motion of test particles.
%
%
The other possibility for obtaining additional minima of the effective potential ${\cal U}_{eff}$ could be realized by introducing a new wave function
$\Psi(x^{M}) = {\cal R}(x^{M})$ $\exp(i \, S(x^{M}))$ for a ``quasar" system,  which fulfills the Klein-Gordon equation (similarly as in \cite{Dziekuje-za-neutron}), where ${\cal R}(x^{M})$ and $S(x^{M})$ are two real field functions. Thus $S(x^{M})$ fulfills the equation which is formally similar to the Hamilton-Jacobi
one, (\ref{row_49}),
except for the modification of the effective potential caused by a potential of the Bohm's type (compare \cite{Davis-Johns,Bohm}). ({\it As Bohm argued,  $S(x^{M})$ is in this case a general integral and not
the complete one.}) \\
\\
{\footnotesize
{\bf Remark}:
{\it
The self-consistent gravitational coupling of the matter field $\Psi(x^{M})$
(included into the total fields configuration)
with the metric tensor causes the breaking of the scale invariance of the action (\ref{row_dzialanie-EH-fi})-(\ref{row_2}), connected with  scaling of $A$ (see
Section~\ref{Scale invariance}).
Thus, due to the Einstein equations, a relation between the four-dimen\-sional mass $m_{4}$ and the parameter $A$ appears whose  value is fixed  in this way (compare \cite{Dziekuje-za-neutron}).}
}
\\
\\
An analysis of these two possibilities will be presented in
following papers.

Next, it would be of observational importance to examine e.g. the apparent surface brightness ${\cal A}$ of such sources as galaxies.
Let $d_{L}$ be the (bolometric) luminosity distance and $d_{\cal A}$ the angular size distance of a galaxy. Then ${\cal A}$ is the quotient of the total flux received by the observer, which behaves like $d_{L}^{-2}$, to the angular area of the galaxy that is seen by the observer,  which goes as $d_{\cal A}^{-2}$, i.e. ${\cal A} \propto (d_{\cal A}/d_{L})^{2}$. Thus, it would be easy to notice that ${\cal A}$ is the function of both the (combined) redshift $z$, (\ref{row_z3}), and the real distance ${\rm R}$ from the source to the observer. \\
\\
{\footnotesize
{\bf Remark}:
{\it Because
of the integral form  (\ref{row_32}) of the physical radial distance of the source from the center of the background gravito-dilatonic field configuration, the apparent surface brightness ${\cal A}$ cannot be written
as a simple analytical expression.}
}\\
\\
Simultaneously, it would also be of interest to examine interactions of the dilatonic-like centers and their distribution $\rho(\varphi)$ in the universe in order to understand its state. The analysis of ${\cal A}$ as the function of $z$ and  ${\rm R}$, which depends on $\rho(\varphi)$, will be discussed in
following papers.\\
Finally, the observations  mentioned  above
\cite{Burbidge} are still growing in number \cite{ostatnie}. These and others \cite{large-quasar-group}, like e.g. the helium abundance in the blue hook stars \cite{blue-hook-stars}, have no
reasonable explanation within the standard interpretation that clai\-ms e.g. that
the Hubble expansion is responsible for the redshifts of galaxies or within the theories with the continuous self-creation of matter and the variable mass hypothesis. \\
There is certainly a need for new models; however, seeking them among the evolutionary ones is not the
proper way.


\section*{Acknowledgments}
This work has been supported by L.J.Ch.. \\
Thanks to Marek Biesiada for the discussions.
This work has been also supported by the Department of Field
Theory and Particle Physics, Institute of Physics, University of
Silesia.

\newpage

\appendix

\section{\label{noncosm-app}The solution for $A < 0$}

If the parameter $A < 0$, then Eqs.(\ref{row_20})-(\ref{varphi
solution}) are valid only when $r > |A|$. Likewise, in Section~\ref{prop}
we write the temporal and radial components of the metric $g_{MN}$
and the internal ``radius'' $\varrho (r)$ (see
Eqs.(\ref{row_21})~and~(\ref{row_26}))
\begin{eqnarray}
\label{row_36}
\left\{
\begin{array}{lll}
g_{tt} = \frac{r}{r - |A|} \; &  &  \\
g_{rr} = - \frac{r}{r - |A|} \; &  &  \\
\varrho (r) = d_{out} \sqrt{\frac{r - |A|}{r}} \; \; . &  &
\end{array}
\right.
\end{eqnarray}
As in Section~\ref{general}, we have $g_{tt} \rightarrow 1$ for
$r \rightarrow \infty$ (see Eq.(\ref{row_31})). But, comparing
the gravitational potential $g_{tt} = \frac{r}{r - |A|} \approx 1
+ \frac{|A|}{r}$ for $r\gg |A|$ with
$g_{tt} = 1 - \frac{G}{c^{2}} \frac{M}{r} $,
which is the one induced by a mass $M$ in the Newtonian limit, we notice that the gravitational potential $g_{tt} $ (see
Eq.(\ref{row_36})) is a repulsive one, contrary to the case for $A > 0$.  \\
The formulae for the scalar curvature ${\cal R}$ (see
Eq.(\ref{row_24})) and dilatonic field $\varphi$ Eq.(\ref{varphi
solution}) now have the form
\begin{eqnarray}
\label{row_38}
{\cal R} = \frac{A^{2}}{2 r^{3} (r - |A|)} \;
\end{eqnarray}
\begin{eqnarray}  \label{row_39}
\varphi (r) = \pm \, \sqrt{\frac{1}{2 \kappa_{6}}} \: \ln(\frac{r}{r - |A|})
\; ,
\end{eqnarray}
where $d_{out} $ is the constant. \\
From (\ref{row_36}) we see that the metric tensor becomes singular at $r = |A|$. However, its determinant $g$ (see Eq.(\ref{row_27})) remains well defined. Nevertheless,
we notice that because at $r = |A|$ the curvature scalar ${\cal R}$ becomes singular and for $r \leq |A|$ the field $\varphi$ is not defined, hence this metric singularity is a genuine one and a system is well posed  only for $r > |A|$. \\
Thus, from Eqs.(\ref{row_38})-(\ref{row_39}), it follows that the physical radial distance can be calculated outside the surface $r = |A|$ only and its value, e.g. from the radius
$|A|$ to the radius $r > |A|$, is equal to
\begin{eqnarray}  \label{row_37}
\!\!\!\!\!\!\!\!\!\!\!\!\!\!
\!\!\!\!\!\!\!\!\!\!\!\!\!\!
& & r_{l-A} = \int_{|A|}^{r} dr \, \sqrt{-
g_{rr}} \\
\!\!\!\!\!\!\!\!\!\!\!\!\!\!
\!\!\!\!\!\!\!\!\!\!\!\!\!\!
& &
= \sqrt{\frac{r}{ r - |A|
}} \, (r - |A|) +  \frac{1}{2} \, |A| \, \ln \left( \frac{ - |A| + 2 r + 2
\, ( r - |A| ) \, \sqrt{\frac{r}{r - |A|}}}{|A|} \right)
> r - |A|. \nonumber
\end{eqnarray}
The region $r > |A|$ is the only one where a frequency shift can be observed. Using Eq.(\ref{row_36}) we can rewrite Eq.(\ref{row_44}) as follows
\begin{eqnarray}  \label{row_47}
\frac{\omega_{obs}}{\omega_{\sigma}} = \frac{\sqrt{\frac{r_{\sigma}}{r_{\sigma} - |A|}}}{
\sqrt{\frac{r_{obs}}{r_{obs} - |A|}}} \; \; .
\end{eqnarray}
Like before, we take for simplicity's sake the limit when the observer is
in infinity. So we get (see Figure~A1)
\begin{eqnarray}  \label{row_48}
\!\!\! \frac{\omega_{obs}}{\omega_{\sigma}} = \sqrt{\frac{r^{w}_{\sigma}}{r^{w}_{\sigma} - 1}} \; ,
\;\; {\rm where} \; \; r^{w}_{\sigma} = \frac{r_{\sigma}}{|A|} \;\; {\rm and} \;\; r_{\sigma} > |A|  .
\end{eqnarray}
\begin{figure}[top]
\begin{center}
\subfigure{\includegraphics[angle=0,width=120mm]{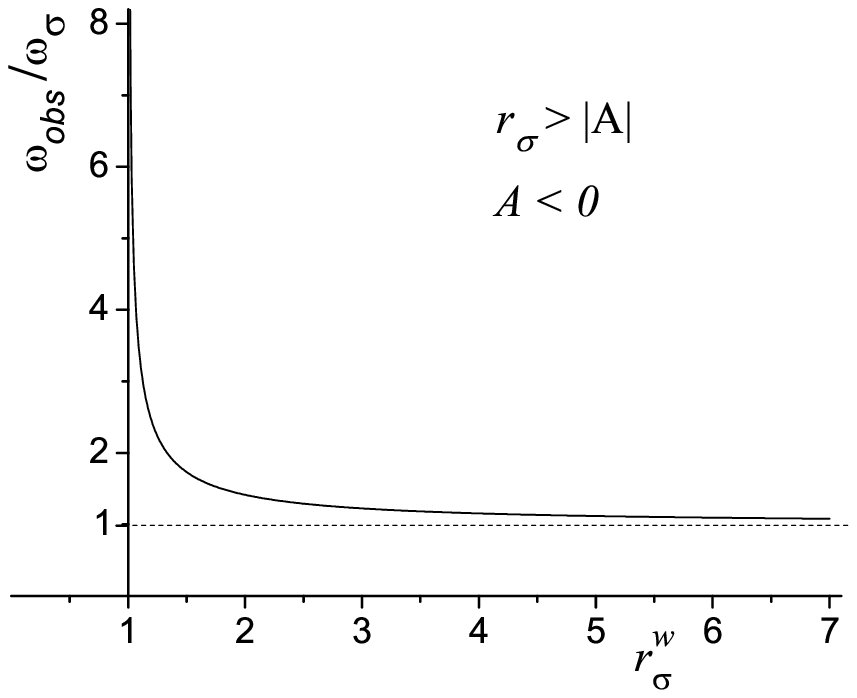}}
\end{center}
\vspace*{8pt}
\caption{
The ratio of the frequency $\omega_{obs}$ of the photon which
reaches the observer (which is at infinity) to the frequency $\omega_{\sigma}$ of the photon
emitted from the source at $r_{\sigma} > |A|$ ($A < O$) as a function of the relative radius
$r^{w}_{\sigma} = \frac{r_{\sigma}}{|A|}$.
\label{f9}
}
\end{figure}\\
We obtained the result that the nearer the source was to the surface given by the equation $r = |A| $ the more the emitted photon which reaches the observer would be blueshifted. This property, caused (in this case) by  the repulsive character of the gravitational potential, is the reason why
this solution was omitted in the main text, as to date it is rather unnoticed in observations. But, because it formally provides a solution to the model, some of its properties have been reproduced  here.

\section{Pericenter shift}
\label{pericenter}

Let us consider the shift
of the point with the minimal $r_{s}$, which is the pericenter
for a test particle elliptically orbiting a central mass $M$
in the spherically symmetric time-independent metric.
In Section~\ref{prop} we pointed out that in the expansion of the metric
to the first order in $\frac{A}{r}$,
the parameter $A/2$ has similar dynamical consequences as the mass $M$.
The approximate post-Newtonian expression for $g_{tt}$ and $g_{rr}$
of the general metric generated by a central mass $M$ can
be written as  follows \cite{Ciufolini-Wheeler}
\begin{eqnarray}  \label{row_postgii}
g_{tt}  & \cong &  \left( 1 - \frac{2 M}{r} +
2 (\beta - \gamma) \frac{M^{2}}{r^{2}} \right) \; \nonumber \\
g_{rr}  & \cong & - \left( 1 + 2 \gamma \frac{M}{r} \right) \; .
\end{eqnarray}
At the lowest order of $\frac{M}{r}$ beyond the Newtonian theory,
an orbit undergoes a pericenter shift \cite{Ciufolini-Wheeler,Misner-Thorne-Wheeler}
given by
\begin{eqnarray}  \label{row_delfi-semi}
\Delta\Phi = \frac{6 \pi M}{a (1 - e^{2})} \times
\frac{(2 + 2 \gamma - \beta)}{3} \; ,
\end{eqnarray}
where $a$ is the semi-major axis of the Newtonian ellipse and $e$ is the eccentricity ($e < 1$ for an ellipse and $e=0$ for the circle). \\
The expression (\ref{row_delfi-semi}) is useful for testing different metric theories of gravity with different field equations.
%
%
For example if we substitute $\gamma \equiv \beta = 1$, corresponding to the general relativity, we get
\begin{eqnarray}  \label{row_delfi-gen-rel}
\Delta\Phi = \frac{6 \pi M}{a (1 - e^{2})} \; .
\end{eqnarray}
Now, let us use the formula given by Eq.(\ref{row_delfi-semi}) for the pericenter shift of a test particle orbiting the center of the dilatonic field $\varphi$, (\ref{varphi solution}), with
the metric tensor given by Eq.(\ref{row_26}) coupled to this field. Neglecting the influence of the internal
two-dimensional space (which is reasonable for $r >> A$) and
expanding $g_{tt}$ and $g_{rr}$ given by Eq.(\ref{row_26})
to the second and first order in $\frac{A}{r}$, respectively, we obtain \begin{eqnarray}
\label{expanssion of g}
g_{tt} \approx 1-\frac{A}{r} + \frac{A^2}{r^2} \;\;\; {\rm and} \;\;\; g_{rr}  \approx  - (1 - \frac{A}{r})  \; .
\end{eqnarray}
Comparing (\ref{expanssion of g}) with the approximate post-Newtonian result (\ref{row_postgii}),
for $M:= A/2$ we get the following approximate values of the post-Newtonian parameters in our model
\begin{eqnarray}
\label{PPN-dilaton}
\gamma = - 1 \;\;\;\;\; {\rm and} \;\;\;\;\; \beta = 1 \; , \;\;\;\;\;
{\rm for} \;\;\; r >> A \; .
\end{eqnarray}
Hence, from Eq.(\ref{row_delfi-semi}) we finally obtain
\begin{eqnarray}  \label{row_delfi-dilaton}
\Delta\Phi = - \, \frac{\pi A}{a (1 - e^{2})} \; .
\end{eqnarray}
Thus, e.g. for a test particle on the orbit with the radius $r_{s} = a$, where $a/A = 100$ (see Table~1
for values of $A$),
%
%
the pericenter shift is equal to  $(-1^{\circ} 48^{'}/_{revolution})$ and $(-9^{\circ} 28^{'}/_{revolution})$ for $e=0$ and $e=0.9$, respectively.
In the post-Newtonian approximation used, the negative sign is characteristic for the influence of the dilatonic field which leads to a retrograde pericenter shift.

\newpage



\end{document}